\documentclass[preprint]{emulateapj}

\mathchardef\mhyphen="2D

\usepackage{subfigure}
\usepackage{graphicx}
\usepackage{amsmath}

\received{}
\revised{}
\accepted{}

\shortauthors{Boyer et al.}
\shorttitle{``The Dust Budget of the SMC''}

\begin{document}

\title{The Dust Budget of the SMC: Are AGB Stars the
  Primary Dust Source at Low Metallicity?}

\author{M.~L.~Boyer\altaffilmark{1},
  S.~Srinivasan\altaffilmark{2},
  D.~Riebel\altaffilmark{3},
  I.~McDonald\altaffilmark{4},
  J.~Th.~van~Loon\altaffilmark{5},
  G.~C.~Clayton\altaffilmark{6},
  K.~D.~Gordon\altaffilmark{1},
  M.~Meixner\altaffilmark{1},
  B.~A.~Sargent\altaffilmark{1},
  G.~C.~Sloan\altaffilmark{7}
  }
  \altaffiltext{1}{STScI, 3700 San Martin Drive, Baltimore, MD 21218 USA; mboyer@stsci.edu}
  \altaffiltext{2}{UPMC-CNRS UMR7095, Institut d'Astrophysique de Paris, F-75014 Paris, France}
  \altaffiltext{3}{Department of Physics and Astronomy, The Johns Hopkins University, Homewood Campus,
  Baltimore, MD 21218 USA}
  \altaffiltext{4}{Jodrell Bank Centre for Astrophysics, Alan Turing Building, University of Manchester, M13 9PL, UK}
  \altaffiltext{5}{Astrophysics Group, Lennard-Jones Laboratories,
  Keele University, Staffordshire ST5 5BG, UK}
  \altaffiltext{6}{Louisiana State University, Department of Physics \& Astronomy, 233-A Nicholson Hall, Tower Dr., Baton Rouge, LA 70803-4001, USA}
  \altaffiltext{7}{Astronomy Department, Cornell University, Ithaca, NY 14853-6801, USA}

\begin{abstract}
  We estimate the total dust input from the cool evolved stars in the
  Small Magellanic Cloud (SMC), using the 8-\micron\ excess emission
  as a proxy for the dust-production rate.  We find that Asymptotic
  Giant Branch (AGB) and red supergiant (RSG) stars produce
  $(8.6$--$9.5) \times 10^{-7}~M_\odot~{\rm yr}^{-1}$ of dust,
  depending on the fraction of far-infrared sources that belong to the
  evolved star population (with 10\%--50\% uncertainty in individual
  dust-production rates). RSGs contribute the least ($<$4\%), while
  carbon-rich AGB stars (especially the so-called ``extreme'' AGB
  stars) account for 87\%--89\% of the total dust input from cool
  evolved stars.  We also estimate the dust input from hot stars and
  supernovae (SNe), and find that if SNe produce $10^{-3}~M_\odot$ of
  dust each, then the total SN dust input and AGB input are roughly
  equivalent. We consider several scenarios of SNe dust production and
  destruction and find that the interstellar medium (ISM) dust can be
  accounted for solely by stellar sources if all SNe produce dust in
  the quantities seen around the dustiest examples and
  if most SNe explode in dense regions where much of the ISM dust is
  shielded from the shocks. We find that AGB stars contribute only
  2.1\% of the ISM dust. Without a net positive contribution from SNe
  to the dust budget, this suggests that dust must grow in the ISM or
  be formed by another unknown mechanism.
\end{abstract}


\vfill\eject
\section{INTRODUCTION}
\label{sec:intro}

Dust in galaxies plays an important role in allowing molecular clouds
to cool sufficiently to form stars and planets. Dust is known to form
in the atmospheres of stars or in explosive/eruptive events such as
novae and supernovae (SNe). There is no known mechanism for forming
dust in the interstellar medium (ISM), though it may be possible to
grow dust in the ISM from existing dust grains
\citep[e.g.,][]{dwek98,draine09}. The rate of grain growth within the
ISM can be estimated by knowing the dust-injection rate from stellar
sources and the dust lifetime within a particular galaxy.

Asymptotic Giant Branch (AGB) stars are important dust creators in
galaxies. Recent work, however, has questioned the long-held belief
\citep{gehrz89} that they are the primary dust factories. For the case
of J114816$+$525150, one of the most distant quasars ($z = 6.4$),
\citet{valiante09} argue that AGB stars must be the source of the
observed dust ($2 \times 10^8 M_\odot$), but \citet{dwek11} argue that
SNe can produce the observed dust if the star formation history is
assumed to have long periods of very low star formation rates. SNe are
generally thought incapable of producing enough dust to account for
the amount seen in the ISM \citep[$\lesssim$$0.01~M_\odot$ each;][and
references therein]{andrews11}. Surprisingly, \citet{matsuura11}
recently detected nearly 1~$M_\odot$ of dust around SN1987A
\citep[see also][]{lakicevic11}; the equivalent dust-injection rate
requires $10^7$--$10^{10}$ dusty AGB stars. However, it is unknown
whether SNe have a net positive or negative impact on dust production,
as they are also efficient dust destroyers.

The only way to compare the total dust input from the evolved stars in
a galaxy to the total dust budget is to detect the entire population
of dusty stars at infrared (IR) wavelengths and estimate the dust-injection rate of each. These sorts of global measurements are
difficult in our Galaxy owing to obscuration by the Galactic Plane and
in Local Group dwarf galaxies owing to limited sensitivity and
resolution \citep[cf.][]{boyer09lg}. Only for the Magellanic Clouds
has the appropriate wavelength coverage, sensitivity, spatial
coverage, and resolution been achieved to attempt to derive the total
dust budget of a star-forming galaxy. In the Large Magellanic Cloud
(LMC), \citet{matsuura09} find that AGB stars and SNe combined account
for only 3\% of the ISM dust, though this could be much higher
if other SNe remnants have large dust reservoirs like
that around SN\,1987A.

Here, we estimate the current global dust production from
the entire population of AGB and red supergiant (RSG) stars in the
Small Magellanic Cloud (SMC). With a metallicity that is 2.5--3
times lower than the LMC \citep[e.g.,][]{russell92,luck98}, the SMC is
more representative of high-redshift galaxies. Combined with estimates
of dust production in SNe, we estimate the total dust budget of the
SMC and compare the results to measurements of the ISM dust mass.

\subsection{Estimating Dust-Production Rates in Evolved Stars}
\label{sec:mlr}

The effect of metallicity on the amount and type of dust species
produced by an evolved star is not yet well understood. Previous
studies \citep[e.g.,][]{groenewegen07,matsuura07,sloan08}
find that more metal-rich stars do appear to produce more oxygen-rich
dust. Presumably, this is because O-rich dust production is limited by
the metallicity-dependent availability of oxygen and silicon, while
C-rich dust production is limited by the mostly
metallicity-independent production of carbon by the star
itself. Metal-poor stars are thus more likely to produce C-rich dust,
as any oxygen is quickly tied up in CO molecules after dredge-up,
leaving an excess of carbon. However, there is evidence that even
carbon stars produce less dust at low metallicity
\citep[e.g.,][]{vanloon00,vanloon08b}, possibly due to a lack of
nucleation seeds.

To accurately estimate the total amount of dust production around a
star, detailed radiative transfer modeling of a well-sampled spectral
energy distribution (SED) is necessary. In addition, the inclusion of
an IR spectrum to determine the dust species is ideal. Studies using
this method have been carried out for subsets of the dust-producing
population in several galaxies and clusters spanning a wide range of
metallicity
\citep[e.g.,][]{vanloon05,matsuura07,groenewegen07,boyer09ngc362,groenewegen09dwarfs,groenewegen09mc,lagadec09,mcdonald09,mcdonald11tuc}.
However, it is currently unfeasible to use this method to measure the
global dust input of an $entire$ population of dust-producing stars
since detailed radiative transfer modeling of thousands of individual
stars is a prohibitively long process. One must thus depend on
photometric techniques for estimating the dust input, usually in the
form of an IR color analysis since IR colors generally scale with the
mass-loss rate \citep[e.g.,][]{groenewegen06,sloan08,matsuura09}. In
this work, a full SED from optical to IR wavelengths is available for
each SMC evolved star, allowing us to use the more physical approach
of computing the IR excess over the stellar photosphere as a proxy for
the dust production \citep[cf.][]{srinivasan09} rather than relying on
the IR colors alone.

\section{Data \& Analysis}
\label{sec:data}

The photometric data used in this study are from the {\it Spitzer}
Legacy program ``Surveying the Agents of Galaxy Evolution in the
SMC'', or SAGE-SMC. Images were obtained at 3.6, 4.5, 5.8, 8.0, 24,
70, and 160~\micron\ covering 30 deg$^2$, including the Tail, Wing,
and Bar of the SMC. Optical to near-IR photometry from the Magellanic
Clouds Photometric Survey \citep[MCPS;][]{zaritsky02}, InfraRed Survey
Facility \citep[IRSF;][]{kato07}, and the 2-Micron All Sky Survey
\citep[2MASS;][]{skrutskie06} were matched to the {\it Spitzer}
photometry, and the full catalog is available for download through the
{\it Spitzer} Science
Center\footnotemark\footnotetext{http://data.spitzer.caltech.edu/popular/sage-smc/}.
For details on the content of the catalog and the nature of the
observations, see \citet{gordon11}.

Throughout, we adopt $A_{\rm V} = 0.12$~mag and $E(B-V) = 0.04$~mag
\citep{schlegel98,harris04} to account for interstellar reddening. We
use the extinction law from \citet{glass99} for optical to near-IR
bands. Extinction in the {\it Spitzer} bands \citep{indebetouw05} is
negligible ($A_\lambda = (6$--$8) \times 10^{-3}$~mag, for $\lambda = 3.6$--$8$~\micron). We assume the distance to the SMC is 60~kpc
\citep{cioni00,keller06}.

AGB and RSG star candidates were selected photometrically from the
catalog, as described by \citet{boyer11}.  AGB candidates were
separated into carbon-rich (C-AGB), oxygen-rich (O-AGB), extreme
(x-AGB), and anomalous O-rich (aO-AGB) sources. The x-AGB sample is
dominated by carbon stars
\citep{vanloon97,vanloon06b,vanloon08b,matsuura09}, though it likely
includes a small number of extreme O-rich sources.  The aO-AGB sources
are a sub-class of the O-AGB candidates with redder $J-[8]$ colors
than the bulk O-AGB population at the same magnitude.  See
\citet{boyer11} for a detailed description of their IR properties.
Since the nature of the aO-AGB sources is unknown and since
\citet{srinivasan09} did not distinguish between the aO- and O-AGB
sources for the LMC analysis, we lump them together for the mass-loss
analysis (Sect.~\ref{sec:mdot}).

 Some cross-contamination between each stellar type and
from young stellar objects (YSOs) and other interlopers is likely, but
we expect it to be minimal \citep[see][]{boyer11}.  Evolved star
candidates showing $F_{24 \micron} > F_{8 \micron}$ are more likely to
include contamination from YSOs or planetary nebulae \citep[Far-IR --
FIR -- objects in][]{boyer11}, so these are considered separately from
the AGB and RSG samples.

\subsection{IR excess}
\label{sec:xses}

We followed \citet{srinivasan09} to estimate the 8-\micron\ excesses
($X_{\rm 8\mhyphen \mu m}$) of the AGB and RSG stars. The 24-\micron\
excess was not considered here because $<$20\% of the O-rich stars are
detected at 24~\micron. See \citet{srinivasan09} for a comparison
between the 8- and 24-\micron\ excesses in the LMC. To compute $X_{\rm
8\mhyphen \mu m}$, we first fit the photospheric emission at optical
and near-IR wavelengths to plane-parallel C-rich COMARCS models from
\citet{Gautschy04} for the C-AGB stars and the spherical O-rich
PHOENIX models from \citet{Hauschildt99} for the O-AGB, aO-AGB, and
RSG stars.  We chose one model for each type of star that best fit
SEDs with little or no dust. The best-fit model photosphere was then
scaled to the $H$-band flux to estimate the IR excess at 8.0~\micron.
Luminosities were determined by a simple trapezoidal integration of the
$U$-band to 24-\micron\ flux.

The x-AGB stars and FIR objects are so heavily obscured in the optical
as to make it impossible to fit the stellar photosphere. For these
sources, we therefore assumed that the flux in the IR is completely
dominated by the IR excess. This is a reasonable assumption since we
expect $<$15\% of the mid-IR flux to come from the stellar photosphere
itself in x-AGB stars (for $J-[3.6] > 3.7$~mag).

The dust mass-loss rate, or dust-production rate (DPR), is expected to
scale with luminosity as $\tau L$ \citep[e.g.,][]{ivezic95}. We
therefore expect the DPR ($\dot{D}$) to increase with luminosity (hence
evolution), provided optical depth ($\tau$) does not decrease with
evolution faster than $L^{-1}$.  Figures~7, 8 and 9 from
\citet{srinivasan09} show that the IR excess also scales with
luminosity for AGB stars, making the excess a good proxy for the
DPR. The relationship between luminosity and $X_{8\mhyphen \micron}$
in the LMC agrees with what we find in the SMC within the
uncertainties.

The 8-\micron\ excess may not scale well with the DPR for the dustiest
O-rich stars that show silicates in absorption at 10~\micron\ rather
than in emission.  We expect these to be rare in the SMC, where the
dust fraction in O-rich stars is low owing to its lower metallicity,
and the smaller population results in fewer stars in this short-lived
evolutionary phase.

To avoid false detections, \citet{srinivasan09} define a
threshold for reliable excesses in terms of data quality -- only
excesses with relative uncertainties less than 1/3 (hereafter,
``$>$3-$\sigma$ excess'') were used in the analysis. We follow this
convention in the current paper. 

\subsection{Derivation of Dust-Production Rates}
\label{sec:rt}

\begin{figure}[h!]
\vbox{
\includegraphics[width=0.48\textwidth]{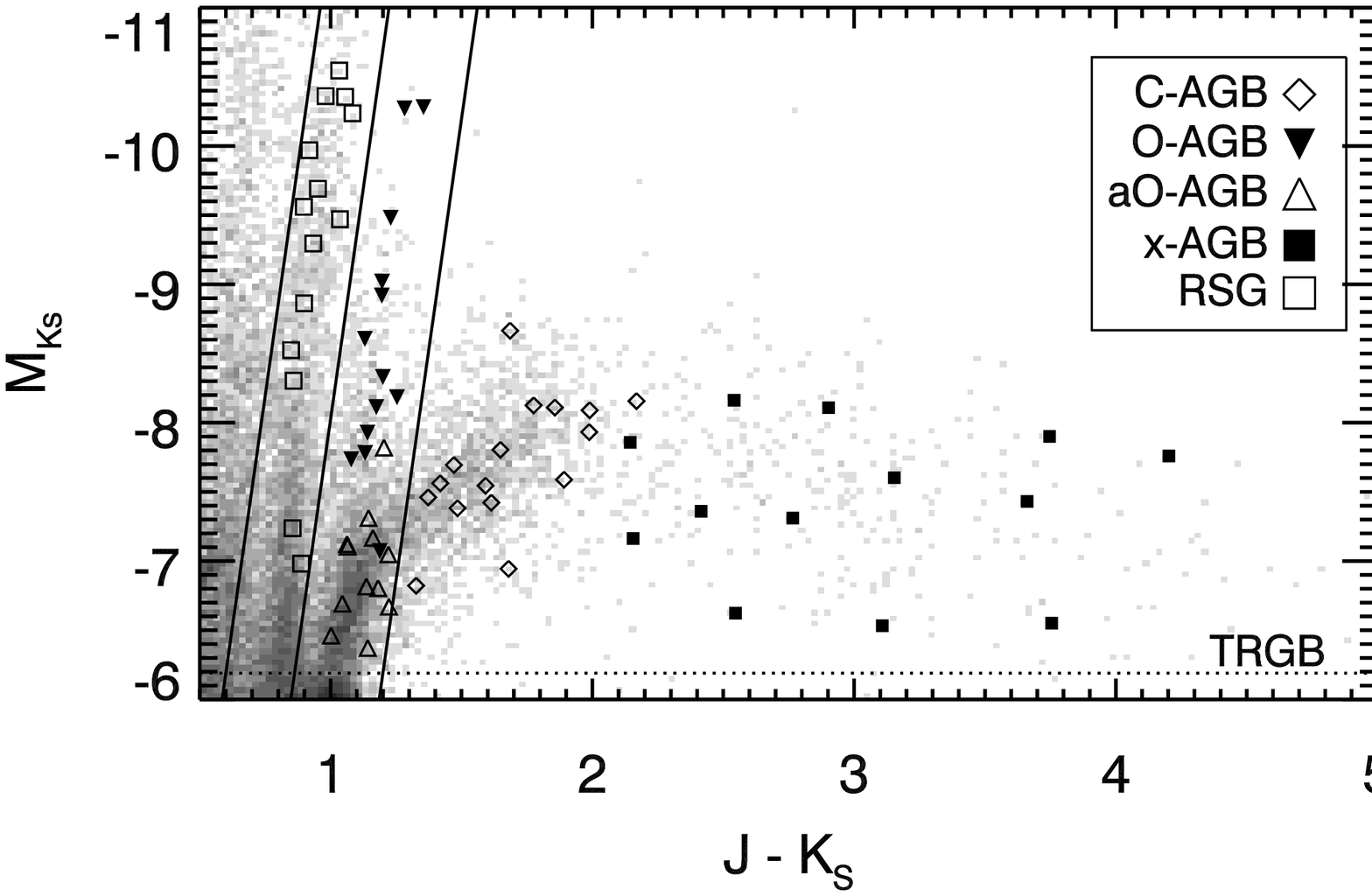}
\includegraphics[width=0.48\textwidth]{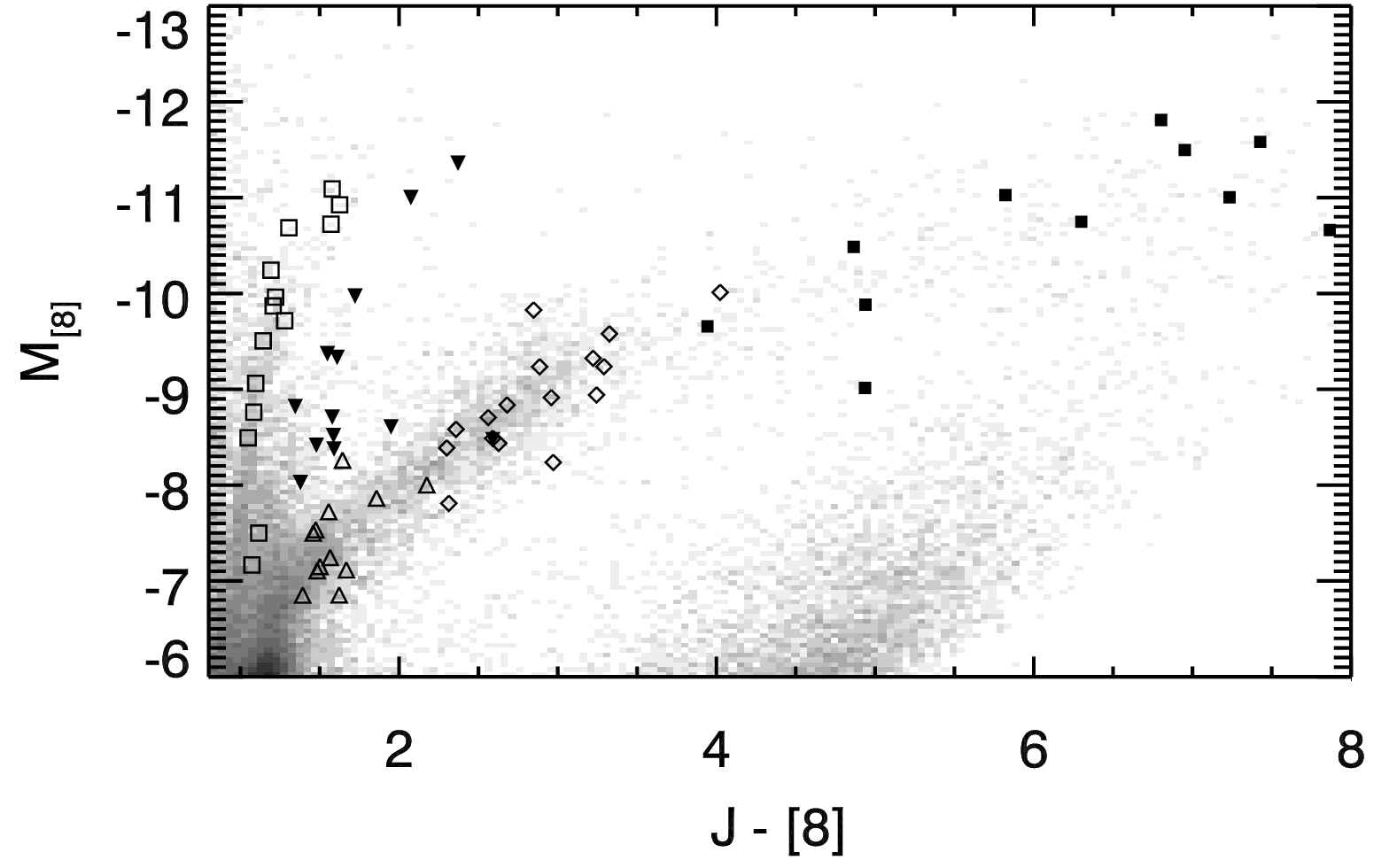}
}
\figcaption{Color-magnitude diagrams (CMD) of evolved stars fit with
the GRAMS models.  The underlying Hess diagram (gray) represents the
full SMC catalog. {\it Upper Panel:} Near-IR CMD. Solid lines show the
approximate division between supergiants, O-rich sources, and C-rich
sources \citep[cf.][]{boyer11}. The dotted line marks the tip of the
Red Giant Branch (TRGB). {\it Lower Panel:} Mid-IR CMD. Stars were
selected for fitting if they have $>$3-$\sigma$ excess and either an
IRS spectrum or $AKARI$ photometry exists to better constrain the
SED. The fitted stars are a representative sample of the SMC evolved
stars. Example SEDs are shown in
Figure~\ref{fig:seds}.\vspace{0.1em}\label{fig:cmd}}
\end{figure}

In order to convert the IR excess to the DPR, we require a
set of stars with known DPRs in the SMC.  \citet{groenewegen09mc}
performed detailed radiative transfer modeling of several stars in the
SMC, but these include mainly x-AGB stars and the dustiest O-AGB
stars.  We therefore select a subset of each type of evolved star and
model their SEDs to determine their DPRs.  We select sources:

\begin{enumerate}

\item that represent SMC AGB and RSG stars over the full IR color space
  (Fig.~\ref{fig:cmd}),

\item that have an 8-\micron\ excess with quality $>$3\,$\sigma$,

\item whose photometric classification as O- or C-rich from
  \citet{boyer11} does not contradict the spectroscopic classification
  from \citet{vanloon08b} or \citet{groenewegen09mc}, and

\item that have additional {\it AKARI} photometry and/or {\it Spitzer}
  InfraRed Spectrograph (IRS) spectra to create a better-constrained
  SED. IRS spectra are from the SMC-Spec {\it Spitzer} program
  (P.I.\ G.\ Sloan); see \citet{kemper10} for a description of IRS data
  reduction. {\it AKARI} photometry is from \citet{ita10}.

\end{enumerate}

\begin{figure}[h!]
\epsscale{1.2} \plotone{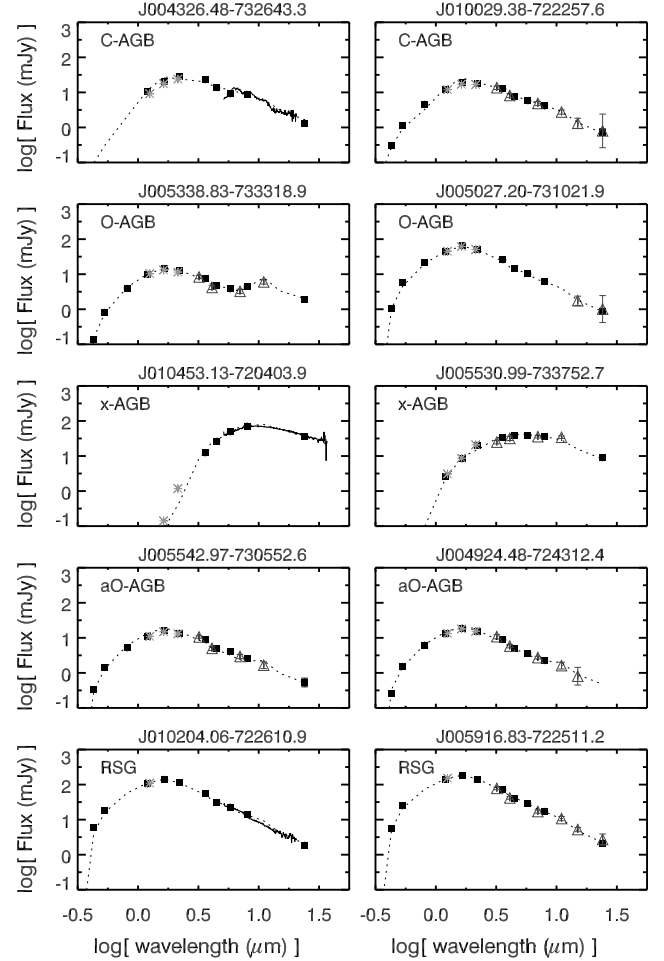} \figcaption{Sample
  SEDs for stars fit with GRAMS models. Photometry from MCPS, 2MASS,
  and {\it Spitzer} is marked by filled squares.  Near-IR photometry
  from the IRSF is marked by asterisks, and $AKARI$ photometry is
  marked by open triangles.  {\it Spitzer} IRS spectra are plotted
  with solid lines, and the best-fit GRAMS model is marked with dotted
  lines connecting synthetic photometry at the observed wavelengths.
  \label{fig:seds}}
\end{figure}

\begin{deluxetable}{lrrr}
\tablewidth{0pc}
\tabletypesize{\normalsize}
\tablecolumns{4}
\tablecaption{Dust mass-loss rates for stars fit with GRAMS models\label{tab:grams}}

\tablehead{\colhead{{\it Spitzer}
    Designation\tablenotemark{a}}&\colhead{Name}&\colhead{Type}&\colhead{log($\dot{D}$)}\\&&&\colhead{($M_\odot$/yr)}} 

\startdata
J003656.74$-$722517.4&    MSX SMC 091& x-AGB& $-$9.0\\
J004650.78$-$714739.2&    MSX SMC 200& x-AGB& $-$8.8\\
J004859.47$-$733538.7&IRAS F00471$-$735& x-AGB& $-$8.2\\
J005000.78$-$724125.5&        \nodata& x-AGB& $-$9.2\\
J005036.97$-$730853.7&        \nodata& x-AGB& $-$9.7\\
J005100.74$-$722518.5&    MSX SMC 163& x-AGB& $-$8.3\\
J005113.57$-$731036.4&        \nodata& x-AGB& $-$8.2\\
J005446.84$-$731337.8&       LEGC 105& x-AGB& $-$9.2\\
J005530.99$-$733752.7&        \nodata& x-AGB& $-$8.9\\
J005548.54$-$724729.3&        \nodata& x-AGB& $-$8.8\\
J005554.63$-$731136.4&        RAW 960& x-AGB& $-$9.4\\
J005718.12$-$724235.2&        \nodata& x-AGB& $-$8.3\\
J010453.13$-$720403.9&  2MASS J010453& x-AGB& $-$7.9\\
J004326.48$-$732643.3&  2MASS J004326& C-AGB& $-$9.8\\
J004912.76$-$732525.5&        \nodata& C-AGB&$-$10.3\\
J004934.01$-$730837.8&        \nodata& C-AGB& $-$9.8\\
J004936.46$-$730357.1&        \nodata& C-AGB&$-$10.4\\
J004956.07$-$724201.3&        \nodata& C-AGB&$-$10.3\\
J005015.40$-$733034.7&        \nodata& C-AGB&$-$10.1\\
J005116.41$-$724256.0&        \nodata& C-AGB&$-$10.0\\
J005127.37$-$724449.3&        \nodata& C-AGB&$-$10.2\\
J005140.46$-$725728.9&    MSX SMC 142& C-AGB& $-$9.5\\
J005607.83$-$731342.6&        \nodata& C-AGB&$-$10.1\\
J005617.52$-$722704.3& S$^3$MC 204803& C-AGB& $-$9.8\\
J005641.02$-$724831.0&        \nodata& C-AGB& $-$9.8\\
J005753.29$-$724343.1&        \nodata& C-AGB&$-$10.4\\
J005918.41$-$722734.2&        \nodata& C-AGB& $-$9.8\\
J005939.00$-$722308.0&        \nodata& C-AGB&$-$10.3\\
J010029.38$-$722257.6&        \nodata& C-AGB&$-$10.4\\
J003201.63$-$732234.7&       HV 11223& O-AGB&$-$10.0\\
J004249.83$-$725511.4&        HV 1366& O-AGB&$-$10.1\\
J004843.11$-$730444.0&        \nodata& O-AGB&$-$10.7\\
J004921.37$-$730327.2&        \nodata& O-AGB&$-$10.0\\
J004950.76$-$724350.0&        \nodata& O-AGB&$-$11.2\\
J004952.05$-$730316.0&        \nodata& O-AGB&$-$10.8\\
J005027.20$-$731021.9&        \nodata& O-AGB&$-$10.8\\
J005338.83$-$733318.9&        \nodata& O-AGB& $-$9.2\\
J005443.62$-$733512.4&        \nodata& O-AGB&$-$10.1\\
J005850.17$-$721835.5&       HV 12149& O-AGB& $-$9.0\\
J010302.42$-$720153.0&        \nodata& O-AGB& $-$8.8\\
J010426.65$-$723440.1&        HV 1963& O-AGB& $-$9.8\\
J004924.48$-$724312.4&        \nodata&aO-AGB&$-$10.3\\
J005000.01$-$730853.1&        \nodata&aO-AGB&$-$10.4\\
J005005.38$-$730500.1&        \nodata&aO-AGB&$-$10.6\\
J005007.31$-$724329.9&        \nodata&aO-AGB&$-$10.4\\
J005051.81$-$733036.8&        \nodata&aO-AGB&$-$10.6\\
J005103.62$-$724612.4&        \nodata&aO-AGB&$-$10.4\\
J005109.94$-$724525.4&        \nodata&aO-AGB&$-$10.5\\
J005401.07$-$733535.6&        \nodata&aO-AGB&$-$10.7\\
J005454.02$-$730806.3&        \nodata&aO-AGB&$-$10.4\\
J005542.97$-$730552.6&        \nodata&aO-AGB&$-$10.2\\
J005855.80$-$723914.5&        \nodata&aO-AGB&$-$10.2\\
J005954.05$-$722219.6&        \nodata&aO-AGB&$-$10.4\\
J004846.36$-$732820.7&        \nodata&   RSG& $-$9.8\\
J004953.78$-$730746.2&        \nodata&   RSG&$-$10.9\\
J005006.34$-$732811.0&    MSX SMC 096&   RSG& $-$9.5\\
J005021.22$-$730609.5&        \nodata&   RSG&$-$10.8\\
J005022.38$-$730755.2&        \nodata&   RSG&$-$11.1\\
J005047.17$-$724257.7&        \nodata&   RSG&$-$10.1\\
J005049.57$-$724154.1&        \nodata&   RSG&$-$10.8\\
J005118.24$-$724324.7&        \nodata&   RSG&$-$10.8\\
J005916.83$-$722511.2&        \nodata&   RSG&$-$11.1\\
J005934.99$-$720406.4&        \nodata&   RSG& $-$9.2\\
J005940.53$-$722055.9&        \nodata&   RSG&$-$11.1\\
J010204.06$-$722610.9&       PMMR 132&   RSG&$-$10.5\\
J010304.34$-$723413.0&       PMMR 141&   RSG&$-$10.5\\
J010315.45$-$724012.2&       PMMR 145&   RSG&$-$10.7

\enddata

\tablenotetext{a}{\ The {\it Spitzer} designation prefix is SSTISAGEMA.}


\end{deluxetable}

To find the DPRs of this subset of evolved stars, we use the Grid of
RSG and AGB ModelS \citep[GRAMS;][]{sargent11,srinivasan11}. These
models cover the full range of stellar and dust properties relevant to
RSG and AGB stars and were developed to reproduce the IR colors of
evolved stars in the LMC. GRAMS fits to the \citet{groenewegen09mc}
carbon stars produce DPRs that are systematically lower by a factor of
2--4 due to a difference in opacities of the amorphous carbon dust
used in the two studies \citep[see Fig. 11 in][]{srinivasan11}. The
GRAMS O-rich fits do not show a systematic offset from the
\citet{groenewegen09mc} values; in fact there is an overall agreement
between DPRs \citep[Fig. 14 in][]{sargent11}. However, it is still
possible that fits for individual stars differ in DPRs by a factor of
up to 6, especially at lower DPRs (this discrepancy is also due to
different optical constants for silicate dust). See
Section~\ref{sec:unc} for a discussion of uncertainties.

We fit the SEDs of the selected SMC sources with GRAMS models using a
simple chi-squared routine. The chi-squared calculation includes
$AKARI$ photometry, where available. IRS spectra are not included in
the chi-squared computation, but are used to confirm by eye whether
the best-fit model is a good match. Sources with poor GRAMS fits are
excluded from the sample, leaving 12 O-AGB, 16 C-AGB, 13 x-AGB, 12
aO-AGB, and 14 RSG stars. We show a sample of these stars in
Figure~\ref{fig:seds} and their DPRs are listed in
Table~\ref{tab:grams}.

GRAMS assumes a wind expansion velocity of 10~km~s$^{-1}$ for all
stars. Assuming that $v_{\rm exp}$ is 10~km~s$^{-1}$ for a star with
$L=30\,000 L_\odot$ in the LMC, we scaled the DPR from GRAMS according
to the following: $\dot{D} \propto L^{0.5}v_{\rm exp}$ and $v_{\rm
exp} \propto L^{0.25}\psi^{-0.5}$, where $\psi$ is the gas-to-dust
ratio \citep[cf.][]{vanloon06}. We assume that $\psi$ scales with
metallicity \citep{vanloon00,marshall04}, $Z_{\rm SMC} = 0.2~Z_\odot$,
and $\psi_\odot = 200$ \citep[e.g.,][]{knapp93,knapp01}, so that
$\psi_{\rm SMC} = 1000$. However, we note that the gas-to-dust ratio
metallicity dependence remains highly uncertain, and may not be the
same for O-rich and C-rich sources. It has been suggested that C-rich
stars may have gas-to-dust ratios similar to Galactic values, even
when in metal-poor environments
\citep[e.g.,][]{habing96,groenewegen07}. For C-rich sources, we
therefore assume $\psi = 200$.

We note that the GRAMS grid does not account for metallic iron dust,
though recent work \citep{mcdonald10, mcdonald11tuc, mcdonald11cen}
shows that it may be an important contributor to dust production,
especially for low-mass, metal-poor stars. The inclusion of metallic
iron may provide substantial changes to the modeled DPRs for the
fainter AGB stars; the direction of this change depends on competing
factors relating to the higher opacity of metallic iron dust. Modeling
of O-rich stars in the globular cluster $\omega$\,Centauri suggests
that the inclusion of metallic iron in a purely dust-driven wind will
increase the DPRs by approximately a factor of 2
\citep{mcdonald09,mcdonald11cen}. However, we expect metallic iron to
be only a minor component of optically thick dust shells, which
dominate the dust production.

\subsection{Extrapolation to the Entire SMC}
\label{sec:mdot}

In Figure~\ref{fig:xses}, we show the DPRs as a function of 8-\micron\
excess for the stars fit in the previous section. The best fits in
Figure~\ref{fig:xses} are as follows:

\begin{figure}[h!]
\epsscale{1.2} \plotone{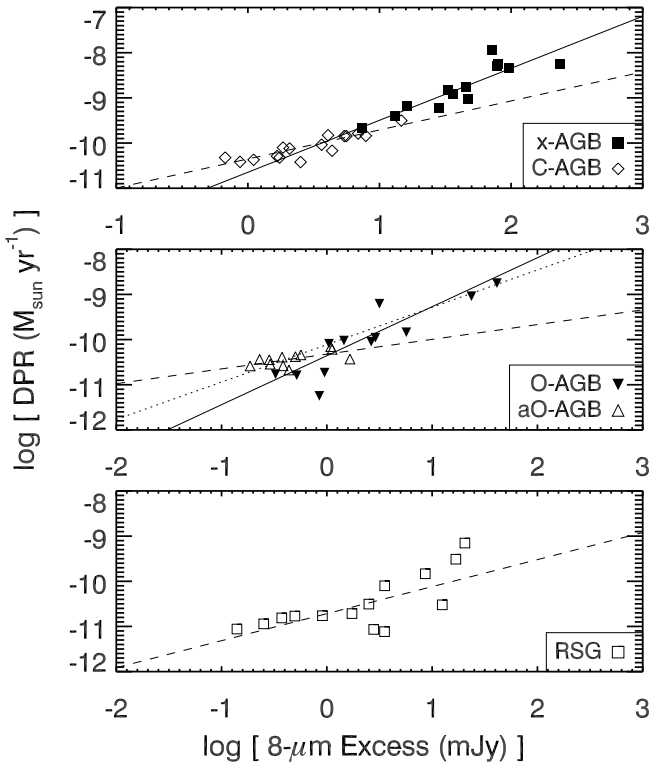}
\figcaption{8-\micron{} excess vs.\ $\dot{D}$ for stars marked in
Fig.~\ref{fig:cmd}. DPRs were determined by applying the GRAMS model
grid (see text).  The solid lines mark the best fits for the filled
symbols, and the dashed lines mark the best fits for the open symbols.
In the middle panel, the dotted line marks the best fit obtained by
combining the O-AGB and aO-AGB stars. We use the combined fit to
derive the DPRs of the entire O-rich AGB sample to
facilitate a direct comparison to the LMC analysis, which did not
distinguish between aO- and O-AGB sources
\citep{srinivasan09}. \label{fig:xses}}
\end{figure}

\begin{equation} \label{eq:mlr}
{\rm log}\,\dot{D}\,(M_\odot~{\rm yr}^{-1}) = A + B~{\rm log}\,X_{\rm 8\mhyphen \mu m}\,{\rm
  (mJy)},
\end{equation}

where

\begin{align*}
&A = -10.7,\ B = 1.2~{\rm (x\mhyphen AGB/FIR),}\\
&A = -10.4,\ B = 0.6~{\rm (C\mhyphen AGB),} \\
&A = -10.4,\ B = 1.1~{\rm (O\mhyphen AGB/aO\mhyphen AGB),} \\
&A = -10.7,\ B = 0.6~{\rm (RSG).}
\end{align*}

The DPR of each evolved star in the SMC is estimated from its
8-\micron\ excess by applying the above $X_{\rm 8\mhyphen \mu
m}$--$\dot{D}$ relationship.  Assuming the FIR objects are dusty
evolved stars, their DPRs are derived using the x-AGB $X_{\rm
8\mhyphen \mu m}$--$\dot{D}$ fit, as most of the FIR sources fall
within the same mid-IR color-magnitude locus as the x-AGB
stars. However, we stress that the majority of the FIR sources are
likely YSOs.

\begin{figure}[h!]
\epsscale{1.2} \plotone{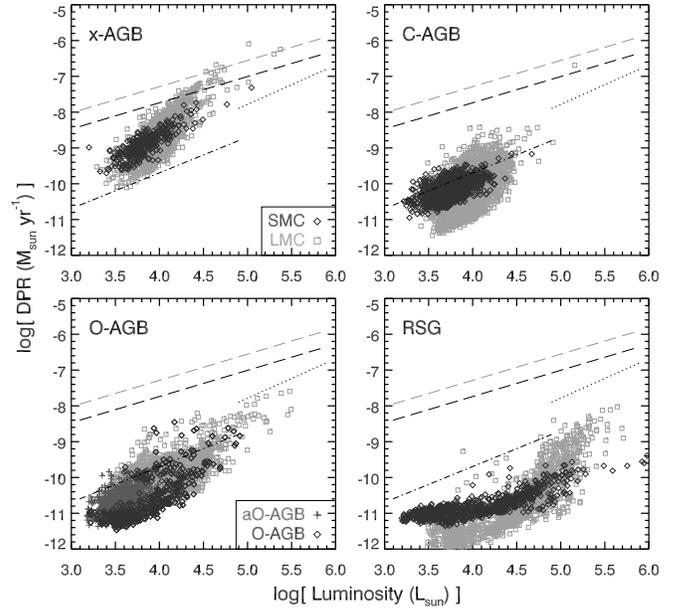} \figcaption{The DPR as
function of luminosity for the SMC and LMC. The long-dashed line
indicates the classical mass loss limit \citep[][$\dot{D} = L(v_{\rm
exp}c)^{-1} \propto L^{0.75}$]{jura84} scaled by the metallicity of
the LMC (grey boxes) and SMC (black diamonds), the dash-dot line indicates the
nuclear consumption rate for the CNO cycle, and the dotted line marks
the nuclear consumption rate for the 3$\alpha$ process.  Both nuclear
consumption rates are directly proportional to $L$. LMC and SMC
sources are plotted with light and dark points, respectively. Only
sources with $>$3-$\sigma$ excess are included. We note that RSGs with
log$[L/L_\odot] \lesssim 4$ are more likely to be AGB stars. \label{fig:lum}}
\end{figure}

\citet{srinivasan09} considered only the x-AGB, C-AGB, and O-AGB
candidates in their analysis of the LMC.  Here, we recompute the DPRs
for the LMC, separating the aO-AGB and FIR sources and including the
RSGs. These DPRs were computed using the $X_{\rm 8\mhyphen \mu
m}$--$\dot{D}$ relationship derived by \citet{srinivasan09}. As with
the SMC, the aO-AGB star DPRs are computed using the O-AGB relation,
and the FIR object DPRs (assuming they are evolved stars) are computed
using the x-AGB relation. The RSGs from \citet{vanloon99},
\citet{vanloon05}, and \citet{groenewegen09mc} are used to calibrate
Equation~\ref{eq:mlr}, resulting in:

\begin{align*}
&A = -11.1,\ B = 1.2~{\rm (LMC\ RSGs).}
\end{align*}

The resulting DPRs for both galaxies are plotted against luminosity in
Figure~\ref{fig:lum}. We see that all of the x-AGB stars lose mass at
a rate that is higher than the nuclear consumption rate (dash-dot and
dotted lines in Fig.~\ref{fig:lum}), implying that mass loss dominates
their subsequent evolution. The same is true for a subset of the C-AGB
and O-AGB stars.

\subsection{The Cumulative Dust-Production Rates}
\label{sec:cum_dpr}

\begin{figure*}
\vbox{
\includegraphics[width=0.5\textwidth]{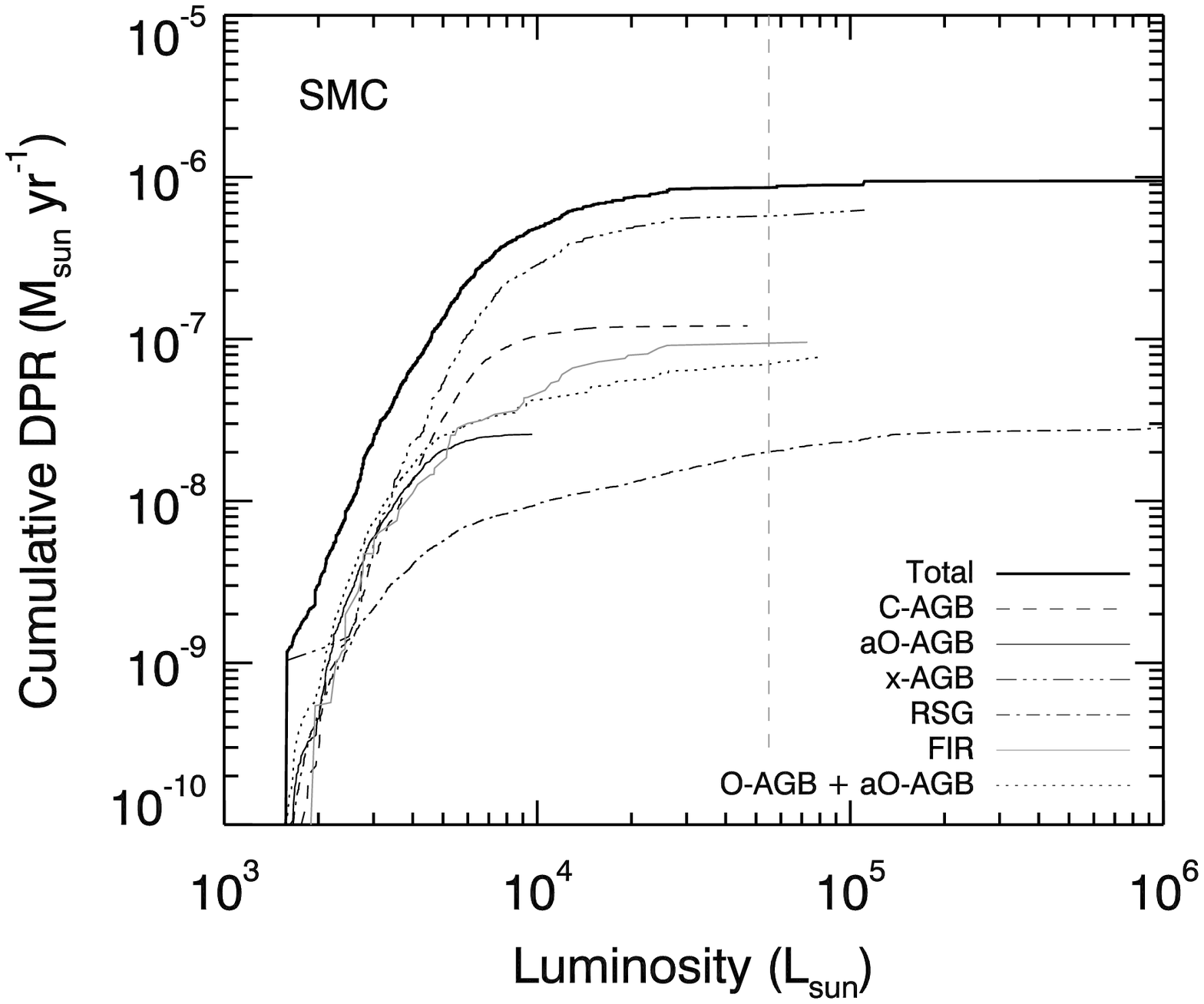}
\includegraphics[width=0.5\textwidth]{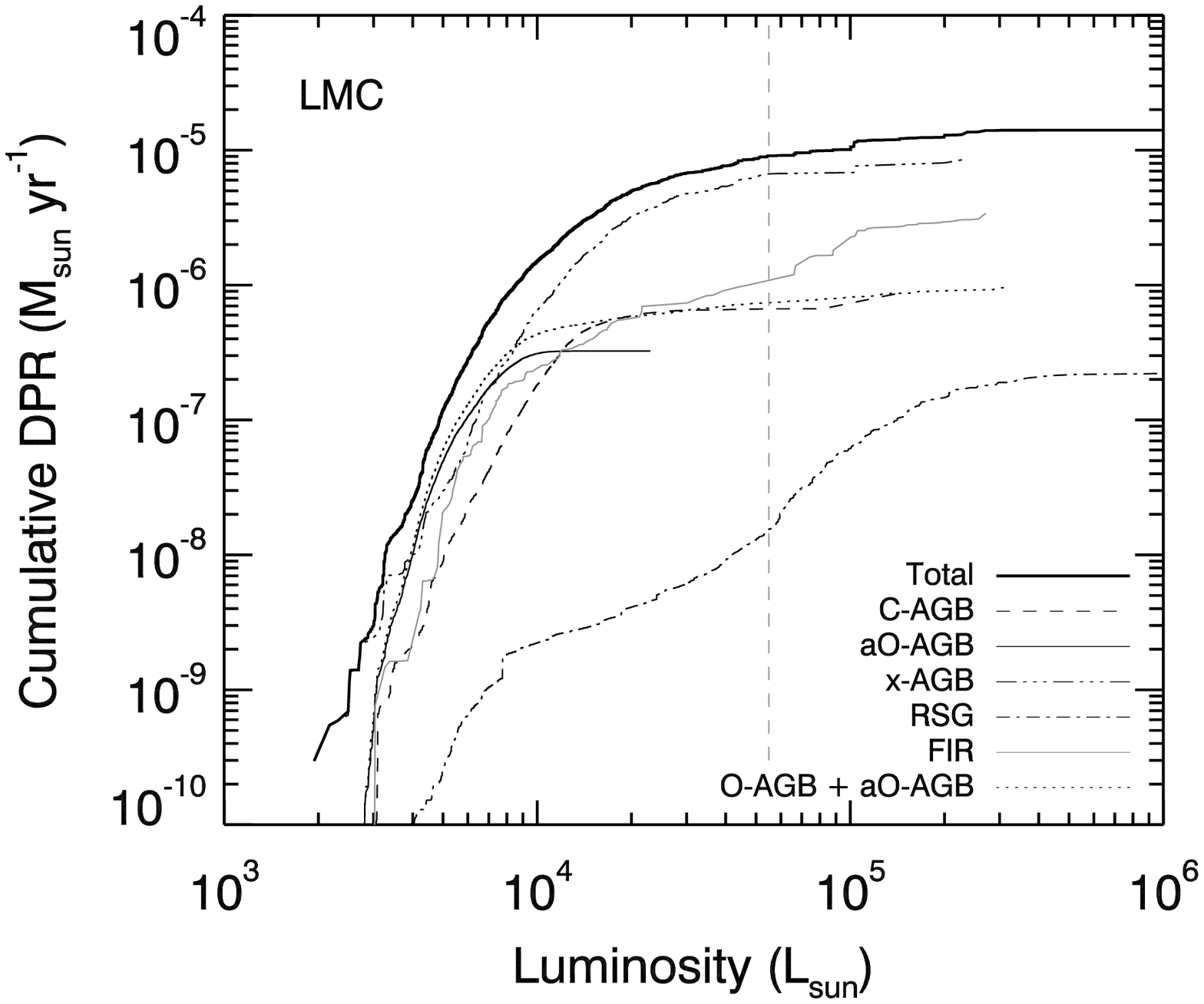}
}
\figcaption{Cumulative DPR vs. luminosity for the SMC
  (left) and LMC (right). The vertical dashed line marks
  the classical AGB limit.\label{fig:mlr}}
\end{figure*}



The cumulative DPRs are shown in Figure~\ref{fig:mlr} and the total dust inputs are listed in
Tables~\ref{tab:mlrs} and \ref{tab:lmc}. The cumulative DPRs in both
galaxies follow similar trends. The x-AGB stars dominate the total DPR
at $L > 4$--$8 \times 10^3~L_\odot$. The O-rich stars dominate the
dust input at lower luminosities, and the regular C-AGB and the O-rich
AGB stars (O-AGB $+$ aO-AGB) contribute similar total DPRs (differing
by a factor of 1 in the LMC and 1.6 in the SMC).  The FIR objects play
a much larger role in the LMC, especially at luminosities higher than
the classical AGB limit. This is likely due to increased contamination
from YSOs in the LMC, which has a higher star formation rate.

We note that stochastics play a role at high luminosity in
Figure~\ref{fig:mlr}, since the brightest evolved
stars are short-lived, and therefore rare. In the SMC, the 10
most extreme x-AGB stars contribute 17\% of the total DPR.

\begin{deluxetable}{lrrrr}
\tablewidth{0pc}
\tabletypesize{\normalsize}
\tablecolumns{5}
\tablecaption{Dust-production rates in the SMC\label{tab:mlrs}}

\tablehead{\colhead{Type}&\colhead{N}&\colhead{$\Sigma \dot{D}$}&\colhead{\% of}&\colhead{$<\dot{D}>$}\\&&\colhead{($M_\odot$/yr)}&\colhead{Total}&\colhead{($M_\odot$/yr)}} 

\startdata
x-AGB  &    313 &  $6.26 \times 10^{-7}$& 65.9 & $2.0 \times 10^{-9\ }$\\
C-AGB  & 1\,559 &  $1.21 \times 10^{-7}$& 12.7 & $7.8 \times 10^{-11}$\\
O-AGB  & 1\,851 &  $0.52 \times 10^{-7}$&  5.5 & $2.8 \times 10^{-11}$\\
aO-AGB & 1\,243 &  $0.26 \times 10^{-7}$&  2.7 & $2.1 \times 10^{-11}$\\
RSG    & 2\,611 &  $0.31 \times 10^{-7}$&  3.3 & $1.2 \times 10^{-11}$\\
FIR    &     50 &  $0.96 \times 10^{-7}$& 10.1 & $1.9 \times 10^{-9\ }$\\
Total  & 7\,627 &  $9.5  \times 10^{-7}$& 100 & $1.2 \times 10^{-10}$\\
Total, no FIR  & 7\,577 & $8.6 \times 10^{-7}$ & 90.5 & $1.1 \times 10^{-10}$  
\enddata



\end{deluxetable}

\begin{deluxetable}{lrrrr}
\tablewidth{0pc}
\tabletypesize{\normalsize}
\tablecolumns{5}
\tablecaption{Dust-production rates in the LMC\label{tab:lmc}}

\tablehead{\colhead{Type}&\colhead{N}&\colhead{$\Sigma \dot{D}$}&\colhead{\% of}&\colhead{$<\dot{D}>$}\\&&\colhead{($M_\odot$/yr)}&\colhead{Total}&\colhead{($M_\odot$/yr)}} 

\startdata
x-AGB  &    886 &  $8.62 \times 10^{-6}$& 61.1 & $9.7 \times 10^{-9\ }$\\
C-AGB  & 5\,190 &  $0.87 \times 10^{-6}$&  6.2 & $1.7 \times 10^{-10}$\\
O-AGB  & 8\,871 &  $0.63 \times 10^{-6}$&  4.5 & $7.1 \times 10^{-11}$\\
aO-AGB & 6\,372 &  $0.32 \times 10^{-6}$&  2.3 & $5.0 \times 10^{-11}$\\
RSG    & 3\,908 &  $0.24 \times 10^{-6}$&  1.7 & $6.1 \times 10^{-11}$\\
FIR    &    179 &  $3.40 \times 10^{-6}$& 24.1 & $1.9 \times 10^{-8\ }$\\
Total  &25\,406 &  $14.1 \times 10^{-6}$& 100 & $5.5 \times 10^{-10}$\\
Total, no FIR & 25\,389 & $10.7 \times 10^{-6}$ & 75.9 & $4.4 \times 10^{-10}$ 
\enddata



\end{deluxetable}

\citet{boyer11} compute the dust input using the $[3.6]-[8]$ color to
compute the DPRs, assuming a dust composition of 85\% amorphous carbon
$+$ 15\% SiC for the carbon stars and 60\% silicates $+$ 40\%
AlO$_{\rm x}$ for the O-rich stars \citep[cf.][]{groenewegen06}.  The
overall trends in the total cumulative DPRs are the same, whether
using the IR excess or the IR color, but the color-derived rates are
4--7 times higher. This discrepancy might be due to an incorrect
assumption about the circumstellar dust
composition. \citet{groenewegen06} shows that if carbon stars have
less SiC and more amorphous carbon, for instance, the color-derived
DPRs will decrease, bringing them closer to the excess-derived
DPRs. SiC abundance is known to decrease with metallicity
\citep[e.g.,][]{lagadec07}, so is expected to be uncommon in SMC
stars. For O-rich stars, a lack of metallic iron or impurities in the
silicates will decrease the 3.6-\micron\ flux and a change in the
shape of the 10-\micron\ silicate feature will affect the
8-\micron\ flux; both will affect how the $[3.6]-[8]$ color scales
with the DPR. Therefore, with the proper assumptions about dust
composition, IR colors may provide DPRs similar to our excess-derived
DPRs.

\subsection{Uncertainties in the Dust-Production Rates}
\label{sec:unc}

The results and analysis presented in this paper are very sensitive to
the DPRs estimated by fitting the observed SEDs to GRAMS
models. Some parameters, such as the outflow velocity, are unknown, so
it is impossible to determine a formal uncertainty in the DPRs. In
this section, we list the known uncertainties associated with the DPR
measurements. Some of these were already discussed in
\citet{sargent11} and \citet{srinivasan11}, but we repeat them here
for completeness.

The current version of the GRAMS grid is computed for spherical dust
shell geometry, assuming a constant DPR throughout the AGB lifetime. A
treatment of the superwind phase is not considered at present. The DPR
is computed from the dust shell inner radius, the size- and
composition-averaged dust grain opacity and the outflow velocity in
the shell \citep[cf.\ Eq.\ 2 in][]{srinivasan10}. When comparing the
GRAMS-predicted DPRs with predictions from other studies, it is
important to account for the differing choices of dust optical
constants; as mentioned in Section~\ref{sec:rt}.

In GRAMS, the outflow velocity is assumed constant and is fixed at
10~km~s$^{-1}$, typical of metal-poor stars \citep{marshall04}. Here,
we scale the DPR to account for the luminosity and the gas-to-dust
ratio (and hence metallicity) dependence of $v_{\rm exp}$
\citep[Sec.~\ref{sec:rt};
e.g.,][]{habing94,ivezic95,vanloon08b,wachter08,lagadec10}.

In performing SED fitting to individual stars using the GRAMS models,
it is necessary to take into account the uncertainties due to
photometric errors as well as stellar variability. This results in a
range of possible model fits, which can be translated to an
uncertainty in the fit parameters, such as luminosity and DPR. A
detailed SED-fitting study (D.Riebel et al., in preparation) finds
that the relative error on the best-fit DPR is 10\%--50\% for the
entire sample of significant mass-losing sources in the LMC. We expect
a similar range of uncertainties in the DPRs used here,
though this uncertainty is statistical and diminishes by integrating
over an ensemble. Also, the relative uncertainty in comparing the SMC
with the LMC is not as large as the absolute uncertainty in each of
their DPRs, owing to the similarity of the data (photometry source,
distance, and morphological type of the galaxies) and the treatment
with the same model grid.

The uncertainties discussed above relate to the determination of DPRs
for individual stars in our sample. The $X_{\rm 8\mhyphen \mu
m}$--$\dot{D}$ relations used in this paper are power-law fits to a
representative sample of sources in the SMC (Fig.~\ref{fig:xses}), and
there is some uncertainty in these fits. For instance, a slope that is
5\% shallower for the $X_{\rm 8\mhyphen \mu m}$--$\dot{D}$ relation
(such that B is 5\% smaller in Eq.~\ref{eq:mlr}) would result in DPRs
10-30\% lower and result in a higher total contribution from stars
with low DPRs. 

An additional source of uncertainty lies in our exclusion of binarity
and the assumption of a spheroidal envelope. In both cases, the
inclination angle could affect the observed 8-\micron\ excess. It is
impossible to determine binarity and/or geometry from IR photometry
alone, but since this analysis includes a large sample of stars
(presumably with no preferred inclination angle with respect to the
line-of-sight), this is not expected to be a large effect.

\section{Discussion}
\label{sec:discussion}

\subsection{Circumstellar Dust at Low Metallicity}
\label{sec:dust}

In Tables~\ref{tab:mlrs} and \ref{tab:lmc}, we show the average DPRs
for stars in the LMC and SMC, and in Figure~\ref{fig:lum}, we show how
the DPR scales with luminosity. Some of the scatter in
Figure~\ref{fig:lum} is due to a spread in the evolutionary phase at
each luminosity. LMC RSGs, which are exclusively O-rich, lose dust at
a rate that is up to 10$\times$ higher than SMC RSGs at the
high-luminosity end (log$[L/L_\odot] \gtrsim 4.5$), consistent with
the findings from several works showing that O-rich dust is more
difficult to form at low metallicity
\citep{vanloon00,vanloon06b,sloan08}. SMC RSGs with low luminosity
appear to have higher DPRs than their LMC counterparts.  This is
likely due to uncertainty in the slope of the $X_{\rm 8\mhyphen \mu
m}$--$\dot{D}$ relationship (Sec.~\ref{sec:unc}).

Previous studies find that the DPR is lower in metal-poor C-rich stars
\citep{vanloon00,vanloon08b}. On average, our findings agree with
this, with $mean$ DPRs 2.2--4.9 times higher in the LMC
(Tables~\ref{tab:mlrs} and \ref{tab:lmc}).  However,
Figure~\ref{fig:lum} shows that the LMC C-rich stars can reach both
high and low DPRs compared to the SMC, and it is clear that metal-poor
C-rich AGB stars can still be prolific dust producers
\citep[e.g.,][]{sloan09}. 

Using a similar treatment of the DPR,
\citet{groenewegen07} find no clear metallicity dependence among
Magellanic Cloud carbon stars. Since our individual DPRs are scaled by
the luminosity and gas-to-dust ratio, they are a factor of 3--9 lower
than those from \citet{groenewegen07}, though this should not change
the relative DPRs in the SMC and LMC, since we use the solar
gas-to-dust ratio ($\psi = 200$) for the carbon stars in both
galaxies.  These results highlight the importance of AGB dust
production at high redshift, where C-rich AGB stars can contribute
strongly to the stellar dust production.

\subsection{Gas Input from Cool Evolved Stars}
\label{sec:gas}

The total current dust-injection rate from the cool evolved stars in
the SMC is $(8.6$--$9.5) \times 10^{-7}~M_\odot~{\rm yr}^{-1}$,
depending on the number of FIR objects which are evolved stars.
Assuming that the gas-to-dust ratio ($\psi$) scales with metallicity
for all stars \citep{vanloon00,marshall04}, then $\psi_{\rm SMC} =
1000$ (see Sec.~\ref{sec:rt}). This yields a total gas return of
$(8.6$--$9.5) \times 10^{-4}~M_\odot~{\rm yr}^{-1}$.  Using Galactic
metallicities for the C-rich objects ($\psi = 200$) and $\psi = 1000$
for the O-rich stars (see Sec.~\ref{sec:rt}) yields a total gas return
of $(2.6$--$2.8) \times 10^{-4}~M_\odot~{\rm yr}^{-1}$, which is close
to the evolved star gas return estimates (derived from dusty evolved
stars) in more metal-poor dwarf irregular galaxies with similar mass
to the SMC such as WLM and IC\,1613 \citep[$\Sigma\dot{D} = (3$--$7)
\times 10^{-4}~M_\odot~{\rm yr}^{-1}$, or
$\approx$$10^{-4}~M_\odot~{\rm yr}^{-1}$ if C-rich sources are
adjusted to solar metallicity;][]{boyer09lg}.

\subsection{Other Dust Sources in the SMC}
\label{sec:other}

We expect additional dust input from SNe and hot massive stars such as
luminous blue variables (LBVs) and Wolf-Rayet (WR) stars, though the
exact amount is highly uncertain.  SNe can destroy dust as well as
produce it.  The SN dust destruction rate is also highly uncertain, and a
large fraction of SN dust may not survive long enough to mix with the
ISM (see Sec.~\ref{sec:survival}). However, several SNe appear to
harbor small amounts of dust.  This includes 0.02~$M_\odot$ of dust
around SN\,2003gd in NGC\,628 \citep{sugerman06,ercolano07},
0.02--0.05~$M_\odot$ around Cas\,A \citep{rho08}, $2.2$--$3.4 \times
10^{-3}~M_\odot$ around SN\,2006jc in UGC\,4905 \citep{sakon09}, $1
\times 10^{-3}~M_\odot$ in the Crab SN remnant \citep{temim06}, and $3
\times 10^{-3}~M_\odot$ around the SMC SN remnant 1E\,0102.2$-$7219
\citep{sandstrom09}.  Surprisingly, recent $Herschel$ far-IR and
ground-based sub-mm observations reveal a massive reservoir of cold
dust around SN\,1987A of 0.4--0.7~$M_\odot$
\citep{matsuura11,lakicevic11}. Based on these observations and
excluding SN\,1987A, we might expect SNe to produce $\sim$$(0.1$--$5)
\times 10^{-2}~M_\odot$ of dust, on average.

\citet{filipovic98} estimate the SN rate ($\tau_{\rm SN}$) in the
SMC to be about 1 every $350 \pm 70$~yr.  \citet{mathewson83} estimate
a more conservative rate of 1 SN every 800~yr. These rates combined
with the observed range of SN dust mass yields a SN dust input of
$(0.1$--$14) \times 10^{-5}~M_\odot~{\rm yr}^{-1}$, the lower limit of
which is comparable to the input from AGB stars.

The SMC is home to 3 LBVs and 5 supergiant B[e] stars that are
detected at 24~\micron\ with MIPS \citep{bonanos10}, which suggests
the presence of circumstellar dust. The dust production by B[e] stars
is not well constrained. Dust inferred from the IR excess may be the
remains of a debris disk left over from the formation of the star
rather than from dust formation in a stellar
outflow. \citet{kastner06} estimate the dust mass in the disks around
two B[e] stars in the LMC, finding $3 \times 10^{-3} M_\odot$ of dust
in the disk around R126.  If formed by the star, this dust must be
ejected in episodic/eruptive events since the mass-loss rate required
is much too high to be sustained over the lifetime of a B[e]
star. Each B[e] star might thus inject $\sim$$10^{-3}~M_\odot$ over
its lifetime. The B[e] phase lasts $\sim$$10^5$~yr, amounting to a
total dust-injection rate of $10^{-7}~M_\odot~{\rm yr}^{-1}$ for 5--10
B[e] stars. This is similar to the dust input from C- and O-AGB stars
in the SMC.

\citet{boyer10} detect a dust-production rate of $2\times 10^{-8}
M_\odot~{\rm yr}^{-1}$ around the LBV R71 in the LMC. If this rate is
typical of dusty LBVs, then their contribution to the total dust
budget is similar to the total DPR of B[e] stars. Dust production in
LBVs is likely episodic, so if we assume that an LBV might lose a few
solar masses of material over its lifetime ($\sim$$10^{4}$~yr) and
assume a gas-to-dust ratio of 1000 for the SMC (see
Section~\ref{sec:gas}), then each LBV might create 0.01~$M_\odot$ of
dust, corresponding to a total DPR of $\sim$$10^{-6}~M_\odot~{\rm
yr}^{-1}$ and matching the input from the cool evolved stars. Galactic
LBVs such as AG\,Carinae and $\eta$\,Carinae create 10 times more dust
than this due to a smaller gas-to-dust ratio
\citep[][]{voors00,smith03}. 

Table~\ref{tab:input} summarizes the estimated dust input from each
dust source. We note that much of the dust from massive stars may be
destroyed in the ensuing shocks when these stars explode as
SNe. Ultimately, the contribution to the total dust budget from
massive stars may be negligible.  We consider the full range of
possibilities in Section~\ref{sec:survival}.

\begin{deluxetable}{lr}
\tablewidth{3.2in}
\tabletypesize{\normalsize}
\tablecolumns{2}
\tablecaption{Dust Sources in the SMC\label{tab:input}}

\tablehead{\colhead{Source}&\colhead{Total Dust Input}\\&\colhead{($M_\odot$/yr)}} 

\startdata
C-rich AGB\tablenotemark{a} & $(7.5$--$8.4) \times 10^{-7}$ \\
O-rich AGB & $7.8 \times 10^{-8}$ \\
RSGs & $3.1 \times 10^{-8}$ \\
LBVs$+$B[e] stars\tablenotemark{b} & $\sim$$10^{-6}$ \\
SNe\tablenotemark{b,c}  & $(0.13$--$14.3) \times 10^{-5}$

\enddata


\tablenotetext{a}{\ The range in dust input from C-rich AGB stars depends on the fraction of FIR stars which are evolved stars.}

\tablenotetext{b}{\ The dust input from LBVs, B[e] stars, and SNe is
highly uncertain. Much of the dust created by massive stars may be
ultimately destroyed by the SNe shock.  See Section~\ref{sec:other}.}

\tablenotetext{c}{\ The range in SNe DPR is computed assuming that SNe input
$\sim$$(0.1$--$5) \times 10^{-2}~M_\odot$ of dust, on average, and
occur at a rate of 350--800 yr (Sec.~\ref{sec:other}).}

\end{deluxetable}

\subsection{Dust Survival in the ISM}
\label{sec:survival}

The dust input into the ISM from stellar sources is listed in
Table~\ref{tab:input}. The dust mass in the SMC ISM inferred from IR
and sub-mm imaging is $(0.29$--$1.1) \times 10^6~M_\odot$
\citep{bot10}. The likelihood that the ISM dust is stellar in origin
depends on the lifetime of dust in the ISM.  Several works find that
SN shocks are the dominant dust-destruction mechanism in the ISM
\citep[e.g.,][]{draine79,jones94}. More than 50\% of a silicate grain
($a = 0.1~\micron$) is returned to the gas phase in shocks with
$v_{\rm s} > 200$~km~s$^{-1}$ \citep{draine79}. \citet{draine09} argue
that a typical SN with energy $E_0 = 10^{51}$~erg in a medium with
density $n_{\rm H} = 1$~cm$^{-3}$ will remain in the Sedov-Taylor
phase until the shock speed drops to 200~km~s$^{-1}$, so we expect a
SN to process $M \approx E_0 / v_{\rm s}^2 = 1260~M_\odot$ of
interstellar material. The dust lifetime within the ISM is then:

\begin{equation} \label{eq:taud}
\tau_{\rm d} = \frac{M_{\rm ISM}^{\rm gas}}{(1260~M_\odot / \tau_{\rm SN})},
\end{equation}

\noindent where $M_{\rm ISM}^{\rm gas}$ (\ion{H}{1} $+$ H$_2$) is $4.5
\times 10^8~M_\odot$ \citep{bolatto11}. \citet{jones96} argue that if
the grains are porous, their lifetimes could be enhanced by a factor
of 3. This, combined with Equation~\ref{eq:taud} and using $\tau_{\rm
SN}$ from Section~\ref{sec:other}, results in an expected dust
lifetime of $\tau_{\rm d} = 0.38$--$0.86$~Gyr, similar to the lifetime
estimated for the Milky Way \citep[$\sim$0.5~Gyr;][]{jones94,
jones96}.

It is possible that many SNe will explode in much denser molecular
clouds, with $n_{\rm H} \sim 10^3$~cm$^{-3}$. In this case, the
Sedov-Taylor phase terminates at higher shock velocities, and the SN
may only process $\sim$170~$M_\odot$ of material, leaving much of the
ISM dust unaffected. If all SNe exploded in dense molecular clouds,
then the dust lifetime increases to $\tau_{\rm d} = 2.8$--$6.7$~Gyr.

If $no$ stellar dust is destroyed by SNe (or other mechanisms such as
grain-grain collisions), then the lifetime of dust in the ISM is
determined by the SFR as $\tau_{\rm d} = M_{\rm ISM}/SFR$, where $SFR$
is the star formation rate \citep[$3.7 \times 10^{-2}~M_\odot~{\rm
yr}^{-1}$;][]{bolatto11}. This yields a lifetime of $\tau_{\rm d}^{\rm
max} = 12$~Gyr, very near the age of the oldest stellar population in
the SMC \citep{noel09}. A dust lifetime of 12~Gyr would suggest that
most dust that has ever been created in the SMC survives in the ISM
today.

In order to compare the dust input to the observed ISM dust mass, we
now consider three cases: 
\begin{itemize}

\item[(A)] massive stars ultimately contribute no dust (it is all
destroyed by the SN shock), such that cool evolved stars are the only
stellar dust sources,

\item[(B)] progenitor dust survives the SN shock and SNe themselves
produce the $lower$ limit of the estimated DPR listed in
Table~\ref{tab:input}, and

\item[(C)] progenitor dust survives the SN shock and SNe themselves
produce the $upper$ limit of the estimated DPR listed in
Table~\ref{tab:input}.

\end{itemize}
\noindent Based on the total dust input and the observed ISM dust
mass, we can estimate the required dust lifetime for each scenario.

For case A, the total dust input is $(8.6$--$9.5) \times
10^{-7}~M_\odot~{\rm yr}^{-1}$ and the dust must therefore survive for
$(3$--$13) \times 10^{11}$~yr, depending on $\tau_{\rm SN}$, to
account for the observed ISM dust mass. This is 2--3 orders of
magnitude longer than the lifetime estimated with
Equation~\ref{eq:taud}, and indeed is longer than a Hubble time. Case
A is therefore an impossible scenario unless the SMC underwent
periods of much higher star formation prior to the era that created
the current population of AGB stars. In 12 regions spanning the
galaxy, \citet{noel09} show that the SFR in the SMC remained constant
within a factor of 3 from the onset of star formation to intermediate
ages, so we do not expect a significantly larger contribution from AGB
stars in the past than is currently observed.

For case B, the total dust input is $(3.3$--$3.4) \times
10^{-6}~M_\odot~{\rm yr}^{-1}$, so the dust lifetime must be
$(9$--$33) \times 10^{11}$~yr.  This lifetime is also much too long,
so the ISM dust cannot be supplied solely by AGB stars and SNe if SNe
produce only $10^{-3}~M_\odot$, on average.

For case C, the total dust input is $1.4 \times 10^{4}~M_\odot~{\rm
yr}^{-1}$ and the dust lifetime must be 2.1--7.9~Gyr, comparable to
the lifetime of porous grains if SNe shockwaves are the dominant
destruction mechanism in the SMC and tend to explode within dense
molecular clouds (Eq.~\ref{eq:taud}: $\tau_{\rm d} =
2.8$--$6.7$~Gyr). SNe that produce $5 \times 10^{-2}~M_\odot$ and
occur every 350~yr can thus feasibly produce the observed ISM dust
mass. We also note that if SNe can produce the amount of dust recently
observed around SN\,1987A (0.4--0.7~$M_\odot$), they can explain the
ISM dust even for the most pessimistic estimates of the dust lifetime.

\subsection{The SMC Dust Budget: Excess Dust in the ISM?}

It is clear that AGB and RSG stars alone cannot account for the
observed ISM dust, even if SNe shocks {\it destroy no dust at
all}. With certain assumptions, SNe and cool evolved stars together can
be made to account for the observed ISM dust mass. This includes
assuming that a high fraction of dust created by SNe and their
progenitors survives and that most SNe explode in a dense medium, such
that most of the surrounding ISM dust is shielded from destruction.
These assumptions are somewhat generous, and we are left with at least
some excess of ISM dust, similar to what is seen in the LMC
\citep{matsuura09}, the Milky Way \citep{dwek98}, and in high-redshift
galaxies. Assuming the dust lifetime derived from
Equation~\ref{eq:taud}, the AGB and RSG stars combined can account for
only up to 2.1\% of the ISM dust. 

Case B in Section~\ref{sec:survival} assumes a conservative estimate
for the SNe dust production. This scenario implies that dust must grow
in the ISM itself, as discussed by \citet{draine09} and references
therein, unless evolved stars produce more dust than is implied by the
mid-IR observations.  Far-IR imaging of a central 2\degr\ $\times$
8\degr\ strip of the LMC with $Herschel$ revealed a strong far-IR
excess ($>200~\micron$) around only one star, the LBV R71
\citep{boyer10}, suggesting that cold dust envelopes are rare around
evolved stars. Cool evolved stars such as AGB and RSG stars are thus
unlikely to be the solution to the missing dust problem. However, if
SN\,1987A is an anomaly or destroys its own dust, the dust input from
cool evolved stars, especially extreme carbon-rich AGB stars, rivals
that from SNe.

Dust accretion in molecular clouds might contribute significantly to
the total dust budget \citep[e.g.,][]{dwek98,zhukovska08}. A chemical
evolution model of the Milky Way derived by \citet{zhukovska08} finds
that the AGB stars dominate the dust input until the metallicity
surpasses $Z \approx 10^{-3}$. The SMC metallicity is $Z = (2 \pm 0.7)
\times 10^{-3}$ \citep{luck98}, so we might expect the dust input from
AGB stars to rival the rate of dust growth in molecular
clouds. Assuming case B in Section~\ref{sec:survival}, we find that
dust grains must grow at a rate of $(0.1$--$3) \times
10^{-3}~M_\odot~{\rm yr}^{-1}$, using conservative estimates for the
dust lifetime ($0.4$--$2.2$~Gyr).  This is similar to the rate of dust
growth in the Milky Way ISM suggested by \citet{jenniskens93}
($10^{-3}~M_\odot~{\rm yr}^{-1}$). The rate of dust accretion in the
ISM may thus be 2--3 orders of magnitude larger than the AGB dust
production.

Alternatively, we must also consider that the ISM recycling timescale
may be long enough such that the current stellar dust input is not
representative of the stellar input during the epoch that created
today's ISM dust. During an increase in star formation, the SNe and
RSGs would contribute more to the dust input. \citet{noel09} find
that the SFR was enhanced 0.2--0.5~Gyr ago in the eastern fields,
perhaps corresponding to a close encounter with the LMC, and resulting
in a larger rate of SNe dust injection during that era.

It is also possible that the ISM dust mass is overestimated. However,
the dust mass assumed here \citep[$(0.29$--$1.1) \times
10^6~M_\odot$;][]{bot10} includes the sub-mm excess, which is
well-constrained by invoking spinning dust grains. If, instead, the
sub-mm excess is due to very cold dust grains, the ISM dust mass would
increase dramatically, exasperating the problem.

\section{Summary of Conclusions}
\label{sec:conclusions}

We estimate the dust-production rates of cool evolved stars in the SMC
using the 8-\micron\ excess and compare the dust input to that from
SNe and to the dust mass in the ISM.  We find that the C-rich AGB
candidates account for 87\%--89\% of the total cool evolved star
dust input. The equivalent fraction in the LMC is 89\%--91\%. The
majority of this dust input comes from the extreme AGB stars.  RSG
stars play a minor role in the dust input, especially below the
classical AGB luminosity limit.

While we can now quantify the dust production in the winds of cool
evolved stars, the SNe dust-production rate (and dust-destruction
rate) remains poorly constrained. It is possible that SNe can account
for all of the ISM dust if they can each produce the upper range of
dust masses observed around SN remnants. If, on the other hand, SNe
can produce only the smaller dust masses inferred from mid-IR
observations of several SN remnants, we expect SNe to contribute
equally to the ISM dust compared to the cool evolved stars. This is
similar to the findings in the LMC, suggesting only a small variance
with metallicity. In this case, an additional dust source is required
and perhaps implies that dust grows efficiently in the ISM.

\acknowledgements This work is supported by NASA via JPL contracts
130827 and 1340964.

\bibliographystyle{astroads}
\bibliography{myrefs}

\begin{thebibliography}{82}
\expandafter\ifx\csname natexlab\endcsname\relax\def\natexlab#1{#1}\fi
\expandafter\ifx\csname href\endcsname\relax
  \def\href#1#2{}\fi
\expandafter\ifx\csname urllinklabel\endcsname\relax
  \def\urllinklabel{[LINK]}\fi
\expandafter\ifx\csname adsurllinklabel\endcsname\relax
  \def\adsurllinklabel{[ADS]}\fi

\bibitem[{{Andrews} {et~al.}(2011){Andrews}, {Sugerman}, {Clayton},
  {Gallagher}, {Barlow}, {Clem}, {Ercolano}, {Fabbri}, {Meixner}, {Otsuka},
  {Welch}, \& {Wesson}}]{andrews11}
{Andrews}, J.~E., {et~al.} 2011, \apj, 731, 47


\bibitem[{{Bolatto} {et~al.}(2011){Bolatto}, {Leroy}, {Jameson}, {Ostriker},
  {Gordon}, {Lawton}, {Stanimirovi{\'c}}, {Israel}, {Madden}, {Hony},
  {Sandstrom}, {Bot}, {Rubio}, {Winkler}, {Roman-Duval}, {van Loon},
  {Oliveira}, \& {Indebetouw}}]{bolatto11}
{Bolatto}, A.~D., {et~al.} 2011, \apj, 741, 12


\bibitem[{{Bonanos} {et~al.}(2010){Bonanos}, {Lennon}, {K{\"o}hlinger}, {van
  Loon}, {Massa}, {Sewilo}, {Evans}, {Panagia}, {Babler}, {Block}, {Bracker},
  {Engelbracht}, {Gordon}, {Hora}, {Indebetouw}, {Meade}, {Meixner}, {Misselt},
  {Robitaille}, {Shiao}, \& {Whitney}}]{bonanos10}
{Bonanos}, A.~Z., {et~al.} 2010, \aj, 140, 416


\bibitem[{{Bot} {et~al.}(2010){Bot}, {Ysard}, {Paradis}, {Bernard}, {Lagache},
  {Israel}, \& {Wall}}]{bot10}
{Bot}, C., {et~al.} 2010, \aap, 523, 20


\bibitem[{{Boyer} {et~al.}(2009{\natexlab{a}}){Boyer}, {McDonald}, {van Loon},
  {Gordon}, {Babler}, {Block}, {Bracker}, {Engelbracht}, {Hora}, {Indebetouw},
  {Meade}, {Meixner}, {Misselt}, {Oliveira}, {Sewilo}, {Shiao}, \&
  {Whitney}}]{boyer09ngc362}
{Boyer}, M.~L., {et~al.} 2009{\natexlab{a}}, \apj, 705, 746


\bibitem[{{Boyer} {et~al.}(2009{\natexlab{b}}){Boyer}, {Skillman}, {van Loon},
  {Gehrz}, \& {Woodward}}]{boyer09lg}
---. 2009{\natexlab{b}}, \apj, 697, 1993


\bibitem[{{Boyer} {et~al.}(2011){Boyer}, {Srinivasan}, {van Loon}, {McDonald},
  {Meixner}, {Zaritsky}, {Gordon}, {Kemper}, {Babler}, {Block}, {Bracker},
  {Engelbracht}, {Hora}, {Indebetouw}, {Meade}, {Misselt}, {Robitaille},
  {Sewi{\l}o}, {Shiao}, \& {Whitney}}]{boyer11}
---. 2011, \aj, 142, 103


\bibitem[{{Boyer} {et~al.}(2010){Boyer}, {van Loon}, {McDonald}, {Gordon},
  {Babler}, {Block}, {Bracker}, {Engelbracht}, {Hora}, {Indebetouw}, {Meade},
  {Meixner}, {Misselt}, {Sewilo}, {Shiao}, \& {Whitney}}]{boyer10}
---. 2010, \apjl, 711, L99


\bibitem[{{Cioni} {et~al.}(2000){Cioni}, {van der Marel}, {Loup}, \&
  {Habing}}]{cioni00}
{Cioni}, M., {et~al.} 2000, \aap, 359, 601


\bibitem[{{Draine}(2009)}]{draine09}
{Draine}, B.~T. 2009, in Astronomical Society of the Pacific Conference Series,
  Vol. 414, Cosmic Dust - Near and Far, ed. {T.~Henning, E.~Gr{\"u}n, \&
  J.~Steinacker}, 453


\bibitem[{{Draine} \& {Salpeter}(1979)}]{draine79}
{Draine}, B.~T. \& {Salpeter}, E.~E. 1979, \apj, 231, 438


\bibitem[{{Dwek}(1998)}]{dwek98}
{Dwek}, E. 1998, \apj, 501, 643


\bibitem[{{Dwek} \& {Cherchneff}(2011)}]{dwek11}
{Dwek}, E. \& {Cherchneff}, I. 2011, \apj, 727, 63


\bibitem[{{Ercolano} {et~al.}(2007){Ercolano}, {Barlow}, \&
  {Sugerman}}]{ercolano07}
{Ercolano}, B., {Barlow}, M.~J., \& {Sugerman}, B.~E.~K. 2007, \mnras, 375, 753


\bibitem[{{Filipovi\'c} {et~al.}(1998){Filipovi\'c}, {Pietsch}, {Haynes},
  {White}, {Jones}, {Wielebinski}, {Klein}, {Dennerl}, {Kahabka}, \&
  {Lazendic}}]{filipovic98}
{Filipovi\'c}, M.~D., {et~al.} 1998, \aaps, 127, 119


\bibitem[{{Gautschy-Loidl} {et~al.}(2004){Gautschy-Loidl}, {H{\"o}fner},
  {J{\o}rgensen}, \& {Hron}}]{Gautschy04}
{Gautschy-Loidl}, R., {et~al.} 2004, \aap, 422, 289


\bibitem[{{Gehrz}(1989)}]{gehrz89}
{Gehrz}, R. 1989, in IAU Symposium, Vol. 135, Interstellar Dust, ed.
  {L.~J.~Allamandola \& A.~G.~G.~M.~Tielens}, 445


\bibitem[{{Glass}(1999)}]{glass99}
{Glass}, I.~S. 1999, {Handbook of Infrared Astronomy}, ed. {Glass, I.~S.}


\bibitem[{{Gordon} {et~al.}(2011){Gordon}, {Meixner}, {Meade}, {Whitney},
  {Engelbracht}, {Bot}, {Boyer}, {Lawton}, {Sewi{\l}o}, {Babler}, {Bernard},
  {Bracker}, {Block}, {Blum}, {Bolatto}, {Bonanos}, {Harris}, {Hora},
  {Indebetouw}, {Misselt}, {Reach}, {Shiao}, {Tielens}, {Carlson},
  {Churchwell}, {Clayton}, {Chen}, {Cohen}, {Fukui}, {Gorjian}, {Hony},
  {Israel}, {Kawamura}, {Kemper}, {Leroy}, {Li}, {Madden}, {Marble},
  {McDonald}, {Mizuno}, {Mizuno}, {Muller}, {Oliveira}, {Olsen}, {Onishi},
  {Paladini}, {Paradis}, {Points}, {Robitaille}, {Rubin}, {Sandstrom}, {Sato},
  {Shibai}, {Simon}, {Smith}, {Srinivasan}, {Vijh}, {Van Dyk}, {van Loon}, \&
  {Zaritsky}}]{gordon11}
{Gordon}, K.~D., {et~al.} 2011, \aj, 142, 102


\bibitem[{{Groenewegen}(2006)}]{groenewegen06}
{Groenewegen}, M.~A.~T. 2006, \aap, 448, 181


\bibitem[{{Groenewegen} {et~al.}(2009{\natexlab{a}}){Groenewegen}, {Lan{\c
  c}on}, \& {Marescaux}}]{groenewegen09dwarfs}
{Groenewegen}, M.~A.~T., {Lan{\c c}on}, A., \& {Marescaux}, M.
  2009{\natexlab{a}}, \aap, 504, 1031


\bibitem[{{Groenewegen} {et~al.}(2009{\natexlab{b}}){Groenewegen}, {Sloan},
  {Soszy{\'n}ski}, \& {Petersen}}]{groenewegen09mc}
{Groenewegen}, M.~A.~T., {et~al.} 2009{\natexlab{b}}, \aap, 506, 1277


\bibitem[{{Groenewegen} {et~al.}(2007){Groenewegen}, {Wood}, {Sloan},
  {Blommaert}, {Cioni}, {Feast}, {Hony}, {Matsuura}, {Menzies}, {Olivier},
  {Vanhollebeke}, {van Loon}, {Whitelock}, {Zijlstra}, {Habing}, \&
  {Lagadec}}]{groenewegen07}
---. 2007, \mnras, 376, 313


\bibitem[{{Habing}(1996)}]{habing96}
{Habing}, H.~J. 1996, \aapr, 7, 97


\bibitem[{{Habing} {et~al.}(1994){Habing}, {Tignon}, \& {Tielens}}]{habing94}
{Habing}, H.~J., {Tignon}, J., \& {Tielens}, A.~G.~G.~M. 1994, \aap, 286, 523


\bibitem[{{Harris} \& {Zaritsky}(2004)}]{harris04}
{Harris}, J. \& {Zaritsky}, D. 2004, \aj, 127, 1531


\bibitem[{{Hauschildt} {et~al.}(1999){Hauschildt}, {Allard}, \&
  {Baron}}]{Hauschildt99}
{Hauschildt}, P.~H., {Allard}, F., \& {Baron}, E. 1999, \apj, 512, 377


\bibitem[{{Indebetouw} {et~al.}(2005){Indebetouw}, {Mathis}, {Babler}, {Meade},
  {Watson}, {Whitney}, {Wolff}, {Wolfire}, {Cohen}, {Bania}, {Benjamin},
  {Clemens}, {Dickey}, {Jackson}, {Kobulnicky}, {Marston}, {Mercer},
  {Stauffer}, {Stolovy}, \& {Churchwell}}]{indebetouw05}
{Indebetouw}, R., {et~al.} 2005, \apj, 619, 931


\bibitem[{{Ita} {et~al.}(2010){Ita}, {Onaka}, {Tanab{\'e}}, {Matsunaga},
  {Matsuura}, {Yamamura}, {Nakada}, {Izumiura}, {Ueta}, {Mito}, {Fukushi}, \&
  {Kato}}]{ita10}
{Ita}, Y., {et~al.} 2010, \pasj, 62, 273


\bibitem[{{Ivezi\'c} \& {Elitzur}(1995)}]{ivezic95}
{Ivezi\'c}, {\v{Z}}. \& {Elitzur}, M. 1995, \apj, 445, 415


\bibitem[{{Jenniskens} {et~al.}(1993){Jenniskens}, {Baratta}, {Kouchi}, {de
  Groot}, {Greenberg}, \& {Strazzulla}}]{jenniskens93}
{Jenniskens}, P., {et~al.} 1993, \aap, 273, 583


\bibitem[{{Jones} {et~al.}(1996){Jones}, {Tielens}, \& {Hollenbach}}]{jones96}
{Jones}, A.~P., {Tielens}, A.~G.~G.~M., \& {Hollenbach}, D.~J. 1996, \apj, 469,
  740


\bibitem[{{Jones} {et~al.}(1994){Jones}, {Tielens}, {Hollenbach}, \&
  {McKee}}]{jones94}
{Jones}, A.~P., {et~al.} 1994, \apj, 433, 797


\bibitem[{{Jura}(1984)}]{jura84}
{Jura}, M. 1984, \apj, 282, 200


\bibitem[{{Kastner} {et~al.}(2006){Kastner}, {Buchanan}, {Sargent}, \&
  {Forrest}}]{kastner06}
{Kastner}, J.~H., {et~al.} 2006, \apjl, 638, L29


\bibitem[{{Kato} {et~al.}(2007){Kato}, {Nagashima}, {Nagayama}, {Kurita},
  {Koerwer}, {Kawai}, {Yamamuro}, {Zenno}, {Nishiyama}, {Baba}, {Kadowaki},
  {Haba}, {Hatano}, {Shimizu}, {Nishimura}, {Nagata}, {Sato}, {Murai},
  {Kawazu}, {Nakajima}, {Nakaya}, {Kandori}, {Kusakabe}, {Ishihara},
  {Kaneyasu}, {Hashimoto}, {Tamura}, {Tanab{\'e}}, {Ita}, {Matsunaga},
  {Nakada}, {Sugitani}, {Wakamatsu}, {Glass}, {Feast}, {Menzies}, {Whitelock},
  {Fourie}, {Stoffels}, {Evans}, \& {Hasegawa}}]{kato07}
{Kato}, D., {et~al.} 2007, \pasj, 59, 615


\bibitem[{{Keller} \& {Wood}(2006)}]{keller06}
{Keller}, S.~C. \& {Wood}, P.~R. 2006, \apj, 642, 834


\bibitem[{{Kemper} {et~al.}(2010){Kemper}, {Woods}, {Antoniou}, {Bernard},
  {Blum}, {Boyer}, {Chan}, {Chen}, {Cohen}, {Dijkstra}, {Engelbracht},
  {Galametz}, {Galliano}, {Gielen}, {Gordon}, {Gorjian}, {Harris}, {Hony},
  {Hora}, {Indebetouw}, {Jones}, {Kawamura}, {Lagadec}, {Lawton}, {Leisenring},
  {Madden}, {Marengo}, {Matsuura}, {McDonald}, {McGuire}, {Meixner}, {Mulia},
  {O'Halloran}, {Oliveira}, {Paladini}, {Paradis}, {Reach}, {Rubin},
  {Sandstrom}, {Sargent}, {Sewilo}, {Shiao}, {Sloan}, {Speck}, {Srinivasan},
  {Szczerba}, {Tielens}, {van Aarle}, {Van Dyk}, {van Loon}, {Van Winckel},
  {Vijh}, {Volk}, {Whitney}, {Wilkins}, \& {Zijlstra}}]{kemper10}
{Kemper}, F., {et~al.} 2010, \pasp, 122, 683


\bibitem[{{Knapp}(2001)}]{knapp01}
{Knapp}, G.~R. 2001, in Astronomical Society of the Pacific Conference Series,
  Vol. 231, Tetons 4: Galactic Structure, Stars and the Interstellar Medium,
  ed. {C.~E.~Woodward, M.~D.~Bicay, \& J.~M.~Shull}, 127


\bibitem[{{Knapp} {et~al.}(1993){Knapp}, {Sandell}, \& {Robson}}]{knapp93}
{Knapp}, G.~R., {Sandell}, G., \& {Robson}, E.~I. 1993, \apjs, 88, 173


\bibitem[{{Lagadec} {et~al.}(2010){Lagadec}, {Zijlstra}, {Mauron}, {Fuller},
  {Josselin}, {Sloan}, \& {Riggs}}]{lagadec10}
{Lagadec}, E., {et~al.} 2010, \mnras, 403, 1331


\bibitem[{{Lagadec} {et~al.}(2007){Lagadec}, {Zijlstra}, {Sloan}, {Matsuura},
  {Wood}, {van Loon}, {Harris}, {Blommaert}, {Hony}, {Groenewegen}, {Feast},
  {Whitelock}, {Menzies}, \& {Cioni}}]{lagadec07}
---. 2007, \mnras, 376, 1270


\bibitem[{{Lagadec} {et~al.}(2009){Lagadec}, {Zijlstra}, {Sloan}, {Wood},
  {Matsuura}, {Bernard-Salas}, {Blommaert}, {Cioni}, {Feast}, {Groenewegen},
  {Hony}, {Menzies}, {van Loon}, \& {Whitelock}}]{lagadec09}
---. 2009, \mnras, 396, 598


\bibitem[{{Laki{\'c}evi{\'c}} {et~al.}(2011){Laki{\'c}evi{\'c}}, {van Loon},
  {Patat}, {Staveley-Smith}, \& {Zanardo}}]{lakicevic11}
{Laki{\'c}evi{\'c}}, M., {et~al.} 2011, \aap, 532, L8


\bibitem[{{Luck} {et~al.}(1998){Luck}, {Moffett}, {Barnes}, \&
  {Gieren}}]{luck98}
{Luck}, R.~E., {et~al.} 1998, \aj, 115, 605


\bibitem[{{Marshall} {et~al.}(2004){Marshall}, {van Loon}, {Matsuura}, {Wood},
  {Zijlstra}, \& {Whitelock}}]{marshall04}
{Marshall}, J.~R., {et~al.} 2004, \mnras, 355, 1348


\bibitem[{{Mathewson} {et~al.}(1983){Mathewson}, {Ford}, {Dopita}, {Tuohy},
  {Long}, \& {Helfand}}]{mathewson83}
{Mathewson}, D.~S., {et~al.} 1983, \apjs, 51, 345


\bibitem[{{Matsuura} {et~al.}(2009){Matsuura}, {Barlow}, {Zijlstra},
  {Whitelock}, {Cioni}, {Groenewegen}, {Volk}, {Kemper}, {Kodama}, {Lagadec},
  {Meixner}, {Sloan}, \& {Srinivasan}}]{matsuura09}
{Matsuura}, M., {et~al.} 2009, \mnras, 396, 918


\bibitem[{{Matsuura} {et~al.}(2011){Matsuura}, {Dwek}, {Meixner}, {Otsuka},
  {Babler}, {Barlow}, {Roman-Duval}, {Engelbracht}, {Sandstrom},
  {Laki{\'c}evi{\'c}}, {van Loon}, {Sonneborn}, {Clayton}, {Long}, {Lundqvist},
  {Nozawa}, {Gordon}, {Hony}, {Panuzzo}, {Okumura}, {Misselt}, {Montiel}, \&
  {Sauvage}}]{matsuura11}
---. 2011, Science, 333, 1258


\bibitem[{{Matsuura} {et~al.}(2007){Matsuura}, {Zijlstra}, {Bernard-Salas},
  {Menzies}, {Sloan}, {Whitelock}, {Wood}, {Cioni}, {Feast}, {Lagadec}, {van
  Loon}, {Groenewegen}, \& {Harris}}]{matsuura07}
---. 2007, \mnras, 382, 1889


\bibitem[{{McDonald} {et~al.}(2011{\natexlab{a}}){McDonald}, {Boyer}, {van
  Loon}, \& {Zijlstra}}]{mcdonald11tuc}
{McDonald}, I., {et~al.} 2011{\natexlab{a}}, \apj, 730, 71


\bibitem[{{McDonald} {et~al.}(2010){McDonald}, {Sloan}, {Zijlstra},
  {Matsunaga}, {Matsuura}, {Kraemer}, {Bernard-Salas}, \&
  {Markwick}}]{mcdonald10}
---. 2010, \apjl, 717, L92


\bibitem[{{McDonald} {et~al.}(2011{\natexlab{b}}){McDonald}, {van Loon},
  {Sloan}, {Dupree}, {Zijlstra}, {Boyer}, {Gehrz}, {Evans}, {Woodward}, \&
  {Johnson}}]{mcdonald11cen}
---. 2011{\natexlab{b}}, \mnras, 417, 20


\bibitem[{{McDonald} {et~al.}(2009){McDonald}, {van Loon}, {Decin}, {Boyer},
  {Dupree}, {Evans}, {Gehrz}, \& {Woodward}}]{mcdonald09}
---. 2009, \mnras, 394, 831


\bibitem[{{No{\"e}l} {et~al.}(2009){No{\"e}l}, {Aparicio}, {Gallart},
  {Hidalgo}, {Costa}, \& {M{\'e}ndez}}]{noel09}
{No{\"e}l}, N.~E.~D., {et~al.} 2009, \apj, 705, 1260


\bibitem[{{Rho} {et~al.}(2008){Rho}, {Kozasa}, {Reach}, {Smith}, {Rudnick},
  {DeLaney}, {Ennis}, {Gomez}, \& {Tappe}}]{rho08}
{Rho}, J., {et~al.} 2008, \apj, 673, 271


\bibitem[{{Russell} \& {Dopita}(1992)}]{russell92}
{Russell}, S.~C. \& {Dopita}, M.~A. 1992, \apj, 384, 508


\bibitem[{{Sakon} {et~al.}(2009){Sakon}, {Onaka}, {Wada}, {Ohyama}, {Kaneda},
  {Ishihara}, {Tanab{\'e}}, {Minezaki}, {Yoshii}, {Tominaga}, {Nomoto},
  {Nozawa}, {Kozasa}, {Tanaka}, {Suzuki}, {Umeda}, {Ohyabu}, {Usui},
  {Matsuhara}, {Nakagawa}, \& {Murakami}}]{sakon09}
{Sakon}, I., {et~al.} 2009, \apj, 692, 546


\bibitem[{{Sandstrom} {et~al.}(2009){Sandstrom}, {Bolatto}, {Stanimirovi{\'c}},
  {van Loon}, \& {Smith}}]{sandstrom09}
{Sandstrom}, K.~M., {et~al.} 2009, \apj, 696, 2138


\bibitem[{{Sargent} {et~al.}(2011){Sargent}, {Srinivasan}, \&
  {Meixner}}]{sargent11}
{Sargent}, B.~A., {Srinivasan}, S., \& {Meixner}, M. 2011, \apj, 728, 93


\bibitem[{{Schlegel} {et~al.}(1998){Schlegel}, {Finkbeiner}, \&
  {Davis}}]{schlegel98}
{Schlegel}, D.~J., {Finkbeiner}, D.~P., \& {Davis}, M. 1998, \apj, 500, 525


\bibitem[{{Skrutskie} {et~al.}(2006){Skrutskie}, {Cutri}, {Stiening},
  {Weinberg}, {Schneider}, {Carpenter}, {Beichman}, {Capps}, {Chester},
  {Elias}, {Huchra}, {Liebert}, {Lonsdale}, {Monet}, {Price}, {Seitzer},
  {Jarrett}, {Kirkpatrick}, {Gizis}, {Howard}, {Evans}, {Fowler}, {Fullmer},
  {Hurt}, {Light}, {Kopan}, {Marsh}, {McCallon}, {Tam}, {Van Dyk}, \&
  {Wheelock}}]{skrutskie06}
{Skrutskie}, M.~F., {et~al.} 2006, \aj, 131, 1163


\bibitem[{{Sloan} {et~al.}(2008){Sloan}, {Kraemer}, {Wood}, {Zijlstra},
  {Bernard-Salas}, {Devost}, \& {Houck}}]{sloan08}
{Sloan}, G.~C., {et~al.} 2008, \apj, 686, 1056


\bibitem[{{Sloan} {et~al.}(2009){Sloan}, {Matsuura}, {Zijlstra}, {Lagadec},
  {Groenewegen}, {Wood}, {Szyszka}, {Bernard-Salas}, \& {van Loon}}]{sloan09}
---. 2009, Science, 323, 353


\bibitem[{{Smith} {et~al.}(2003){Smith}, {Gehrz}, {Hinz}, {Hoffmann}, {Hora},
  {Mamajek}, \& {Meyer}}]{smith03}
{Smith}, N., {et~al.} 2003, \aj, 125, 1458


\bibitem[{{Srinivasan} {et~al.}(2009){Srinivasan}, {Meixner}, {Leitherer},
  {Vijh}, {Volk}, {Blum}, {Babler}, {Block}, {Bracker}, {Cohen}, {Engelbracht},
  {For}, {Gordon}, {Harris}, {Hora}, {Indebetouw}, {Markwick-Kemper}, {Meade},
  {Misselt}, {Sewilo}, \& {Whitney}}]{srinivasan09}
{Srinivasan}, S., {et~al.} 2009, \aj, 137, 4810


\bibitem[{{Srinivasan} {et~al.}(2010){Srinivasan}, {Sargent}, {Matsuura},
  {Meixner}, {Kemper}, {Tielens}, {Volk}, {Speck}, {Woods}, {Gordon},
  {Marengo}, \& {Sloan}}]{srinivasan10}
---. 2010, \aap, 524, A49


\bibitem[{{Srinivasan} {et~al.}(2011){Srinivasan}, {Sargent}, \&
  {Meixner}}]{srinivasan11}
{Srinivasan}, S., {Sargent}, B.~A., \& {Meixner}, M. 2011, \aap, 532, A54


\bibitem[{{Sugerman} {et~al.}(2006){Sugerman}, {Ercolano}, {Barlow}, {Tielens},
  {Clayton}, {Zijlstra}, {Meixner}, {Speck}, {Gledhill}, {Panagia}, {Cohen},
  {Gordon}, {Meyer}, {Fabbri}, {Bowey}, {Welch}, {Regan}, \&
  {Kennicutt}}]{sugerman06}
{Sugerman}, B.~E.~K., {et~al.} 2006, Science, 313, 196


\bibitem[{{Temim} {et~al.}(2006){Temim}, {Gehrz}, {Woodward}, {Roellig},
  {Smith}, {Rudnick}, {Polomski}, {Davidson}, {Yuen}, \& {Onaka}}]{temim06}
{Temim}, T., {et~al.} 2006, \aj, 132, 1610


\bibitem[{{Valiante} {et~al.}(2009){Valiante}, {Schneider}, {Bianchi}, \&
  {Andersen}}]{valiante09}
{Valiante}, R., {et~al.} 2009, \mnras, 397, 1661


\bibitem[{{van Loon}(2000)}]{vanloon00}
{van Loon}, J.~{\relax Th}. 2000, \aap, 354, 125


\bibitem[{{van Loon}(2006)}]{vanloon06b}
{van Loon}, J.~{\relax Th}. 2006, in Astronomical Society of the Pacific
  Conference Series, Vol. 353, Stellar Evolution at Low Metallicity: Mass Loss,
  Explosions, Cosmology, ed. {H.~J.~G.~L.~M.~Lamers, N.~Langer, T.~Nugis, \&
  K.~Annuk}, 211


\bibitem[{{van Loon} {et~al.}(2005){van Loon}, {Cioni}, {Zijlstra}, \&
  {Loup}}]{vanloon05}
{van Loon}, J.~{\relax Th}., {et~al.} 2005, \aap, 438, 273


\bibitem[{{van Loon} {et~al.}(2008){van Loon}, {Cohen}, {Oliveira}, {Matsuura},
  {McDonald}, {Sloan}, {Wood}, \& {Zijlstra}}]{vanloon08b}
---. 2008, \aap, 487, 1055


\bibitem[{{van Loon} {et~al.}(1999){van Loon}, {Groenewegen}, {de Koter},
  {Trams}, {Waters}, {Zijlstra}, {Whitelock}, \& {Loup}}]{vanloon99}
---. 1999, \aap, 351, 559


\bibitem[{{van Loon} {et~al.}(2006){van Loon}, {McDonald}, {Oliveira}, {Evans},
  {Boyer}, {Gehrz}, {Polomski}, \& {Woodward}}]{vanloon06}
---. 2006, \aap, 450, 339


\bibitem[{{van Loon} {et~al.}(1997){van Loon}, {Zijlstra}, {Whitelock},
  {Waters}, {Loup}, \& {Trams}}]{vanloon97}
---. 1997, \aap, 325, 585


\bibitem[{{Voors} {et~al.}(2000){Voors}, {Waters}, {de Koter}, {Bouwman},
  {Morris}, {Barlow}, {Sylvester}, {Trams}, \& {Lamers}}]{voors00}
{Voors}, R.~H.~M., {et~al.} 2000, \aap, 356, 501


\bibitem[{{Wachter} {et~al.}(2008){Wachter}, {Winters}, {Schr{\"o}der}, \&
  {Sedlmayr}}]{wachter08}
{Wachter}, A., {et~al.} 2008, \aap, 486, 497


\bibitem[{{Zaritsky} {et~al.}(2002){Zaritsky}, {Harris}, {Thompson}, {Grebel},
  \& {Massey}}]{zaritsky02}
{Zaritsky}, D., {et~al.} 2002, \aj, 123, 855


\bibitem[{{Zhukovska} {et~al.}(2008){Zhukovska}, {Gail}, \&
  {Trieloff}}]{zhukovska08}
{Zhukovska}, S., {Gail}, H.-P., \& {Trieloff}, M. 2008, \aap, 479, 453


\end{thebibliography}

\end{document}